\documentclass{emulateapj}

\usepackage{graphicx}
\usepackage{psfig}
\usepackage{apjfonts}

\def\cm{\textrm{cm}}

\def\erg{\textrm{ergs}}
\def\kpc{\textrm{kpc}}
\def\pc{\textrm{pc}}
\def\Mpc{\textrm{Mpc}}

\def\ergps{\textrm{ergs}~\textrm{s}^{-1}}

\def\kms{\textrm{km}~\textrm{s}^{-1}}
\def\gcm2{\textrm{g}~\textrm{cm}^{-2}}
\def\ergscm3{\textrm{ergs}~\textrm{s}^{-1}~\textrm{cm}^{-3}}
\def\ergcm3{\textrm{ergs}~\textrm{cm}^{-3}}
\def\gscm2{\textrm{g}~\textrm{s}^{-1}~\textrm{cm}^{-2}}
\def\ergcmK34{\textrm{ergs}~\textrm{cm}^{-3}~\textrm{K}^{-4}}
\def\cms31{\textrm{cm}^{-3}~\textrm{s}^{-1}}
\def\cmg21{\textrm{cm}^{2}~\textrm{g}^{-1}}

\def\phcm2s1{\textrm{photons}~\textrm{cm}^{-2}~\textrm{s}^{-1}}
\def\phFluxUnits{\textrm{ph}~\textrm{cm}^{-2}~\textrm{s}^{-1}}
\def\DiffphFluxUnits{\textrm{ph}~\textrm{cm}^{-2}~\textrm{s}^{-1}~\textrm{GeV}^{-1}}

\def\FUnits{\textrm{ergs}~\textrm{cm}^{-2}~\textrm{s}^{-1}}
\def\eV{\textrm{eV}}
\def\MeV{\textrm{MeV}}
\def\GeV{\textrm{GeV}}
\def\TeV{\textrm{TeV}}
\def\PeV{\textrm{PeV}}

\def\GHz{\textrm{GHz}}

\def\yr{\textrm{yr}}
\def\Myr{\textrm{Myr}}

\def\muGauss{\mu\textrm{G}}

\def\GeVs1cm3{\textrm{GeV}~\textrm{s}^{-1}~\textrm{cm}^{3}}
\def\log{\textrm{log}}
\def\Msun{\textrm{M}_{\sun}}

\newcommand{\mean}[1]{\ensuremath{\langle #1 \rangle}}

\begin{document}

\title{On The GeV \& TeV Detections of the Starburst Galaxies M82 \& NGC 253}

\author{Brian C.~Lacki\altaffilmark{1,2}, Todd A.~Thompson\altaffilmark{1,2,3}, 
Eliot Quataert\altaffilmark{4}, Abraham Loeb\altaffilmark{5},  \& Eli Waxman\altaffilmark{6}}

\altaffiltext{1}{Department of Astronomy
The Ohio State University, Columbus, Ohio 43210, USA}
\altaffiltext{2}{Center for Cosmology \& Astro-Particle Physics,
The Ohio State University, Columbus, Ohio 43210, USA}
\altaffiltext{3}{Alfred P.~Sloan Fellow}
\altaffiltext{4}{Astronomy Department \& Theoretical Astrophysics Center, 601 Campbell Hall, The University of California, Berkeley, CA 94720, USA}
\altaffiltext{5}{Astronomy Department, Harvard University, 60 Garden Street, Cambridge, MA 02138, USA}
\altaffiltext{6}{Physics Faculty, Weizmann Institute, Rehovot 7600, Israel}

\begin{abstract}
The GeV and TeV emission from M82 and NGC 253 observed by Fermi, HESS, and VERITAS constrains the physics of cosmic rays (CRs) in these dense starbursts.  We argue that the $\gamma$-rays are predominantly hadronic in origin, as expected by previous studies.  The measured fluxes imply that pionic losses are efficient for CR protons in both galaxies: we show that a fraction $F_{\rm cal} \approx 0.2 - 0.4$ of the energy injected in high energy primary CR protons is lost to inelastic proton-proton collisions (pion production) before escape, producing $\gamma$-rays, neutrinos, and secondary electrons and positrons.  We discuss the factor $\sim 2$ uncertainties in this estimate, including supernova rate and leptonic contributions to the GeV-TeV emission.  We argue that $\gamma$-ray data on ULIRGs like Arp 220 can test whether M82 and NGC 253 are truly calorimetric, and we present upper limits on Arp 220 from the \emph{Fermi} data.  We show that the observed ratio of the GeV to GHz fluxes of the starbursts suggests that non-synchrotron cooling processes are important for cooling the CR electron/positron population.  We briefly reconsider previous predictions in light of the $\gamma$-ray detections, including the starburst contribution to the $\gamma$-ray background and CR energy densities.  Finally, as a guide for future studies, we list the brightest star-forming galaxies on the sky and present updated predictions for their $\gamma$-ray and neutrino fluxes.
\end{abstract}


\section{Introduction}

M82 and NGC 253 are nearby ($D \approx 2.5 - 4.0\ \Mpc$), prototypical starburst galaxies, each having an intense star-forming region of about $200\ \pc$ radius in the center of a more quiescent disk galaxy.  The starbursts are expected to have high supernova (SN) rates of about $0.03 - 0.3\ \yr^{-1}$.  SN remnants are believed to accelerate primary cosmic ray (CRs) protons and electrons.  The high SN rates in starbursts imply high CR emissivities.  The presence of CR electrons and positrons in these starbursts is inferred from the nonthermal synchrotron radio emission they produce \citep[e.g.,][]{Klein88,Volk89}.  However, most of the CR energy is believed to be in the form of CR protons.  

When high energy CR protons collide with interstellar medium (ISM) nucleons, they create pions, which decay into secondary electrons and positrons, $\gamma$-rays, and neutrinos.  With their dense ISMs ($\mean{n} \approx 100 - 500\ \cm^{-3}$) and possible high CR energy densities \citep[as evinced by the bright radio emission][]{Volk89,Akyuz91,Persic10}, M82 and NGC 253 are predicted to be bright $\gamma$-ray sources (e.g., \citealt{Akyuz91,Sreekumar94,Volk96,Paglione96,Romero03,Domingo05,Thompson07} [TQW]; \citealt{Persic08,deCeaDelPozo09,Rephaeli09,Lacki10} [LTQ]).  As prototypical starbursts, if M82 and NGC 253 are seen in $\gamma$-rays, starbursts in general may be sources of $\gamma$-rays \citep{Pohl94,Torres04a}, with important implications for the diffuse $\gamma$-ray and neutrino backgrounds (e.g., \citealt{Pavlidou02}; \citealt{Loeb06} [LW06]; TQW).  However, the $\gamma$-ray luminosity of starbursts depends not only on the injection rate of CRs, but also on the efficiency of converting CR proton energy into pionic $\gamma$-rays, neutrinos, and secondary electrons and positrons.  In turn, this efficiency depends on the ratio of the timescale for pion production to the escape timescale.  The hypothesis that CR protons in starbursts lose all of their energy to pionic collisions before escaping is called ``proton calorimetry'' \citep[c.f.][]{Pohl94}.\footnote{Here, we consider only CR protons with kinetic energy above the threshold for pion production.}  If proton calorimetry is strongly violated, then M82 and NGC 253 and, by extension, other starbursts could in fact be weak $\gamma$-ray sources.

Although $\gamma$-ray emission from M82 and NGC 253 has been sought for several years with no success (at GeV, \citealt{Cillis05}; and at TeV, \citealt{Aharonian05,Itoh07}), the launch of \emph{Fermi} and the advent of powerful VHE $\gamma$-ray telescopes has led to recent detections of both starbursts at GeV energies (with \emph{Fermi}; \citealt{Abdo10a}) and in VHE $\gamma$-rays (M82 with VERITAS, \citealt{Acciari09}; NGC 253 with HESS, \citealt{Acero09}).  These GeV and TeV detections constrain the cosmic ray (CR) population in these dense star-forming environments.  

In this paper, we discuss the implications of the $\gamma$-ray detections of M82 and NGC 253.  The ratio of the $\gamma$-ray luminosities to the bolometric luminosities informs the question of whether or not these systems are proton calorimeters (TQW).  The emission also has implications for the energy density of CRs in starbursts \citep[e.g.,][]{Akyuz91}.  Finally, combined with the radio emission, the energy losses of CR electrons and positrons are constrained \citep[c.f.,][]{Paglione96,Domingo05,Persic08,deCeaDelPozo09,Rephaeli09}.  Pionic $\gamma$-rays must be accompanied by secondary positrons and electrons; the ratio of the power in these expected electrons and positrons to the observed radio emission informs us of the energy losses of the CR electrons and positrons.  In particular, we derive in \S~\ref{sec:FRCImplications} the expected synchrotron luminosity from the pionic luminosity if synchrotron cooling is the dominant loss process.

In \S\ref{sec:Detections} we describe the detections of M82 and NGC 253 at GeV and TeV energies.  We then interpret the detections as $\gamma$-rays from diffuse CR protons in \S\ref{sec:Interpretation}.  Our interpretation includes comparison of the $\gamma$-ray luminosities of M82 and NGC 253 with their CR luminosities and their IR luminosities (\S\ref{sec:CalorimetryFraction}), and a discussion of the uncertainties in these estimates (\S\ref{sec:Uncertainties}). We find that a fraction 0.4 and 0.2 of luminosity in $\ge \GeV$ CR protons is lost to pion production in M82 and NGC 253, respectively.  We discuss the implications for our estimates mean for proton calorimetry in M82 and NGC 253 are at GeV energies (\S\ref{sec:Calorimetry}).  Other possible sources for the observed $\gamma$-rays are considered in \S\ref{sec:Sources}.  The implications of the detections of M82 and NGC 253 for the detection of other star-forming galaxies, the starburst contribution to the diffuse extragalactic $\gamma$-ray and neutrino backgrounds, the dynamical importance of CRs in starbursts, and for the physics of the FIR-radio correlation are described in \S\ref{sec:Implications}.  We summarize our results in \S\ref{sec:Conclusion}.

\section{$\gamma$-ray Detections}
\label{sec:Detections}

\subsection{\emph{Fermi} and TeV detections}
\label{sec:TeVComparison}
\begin{figure*}
\centerline{\includegraphics[width=9cm]{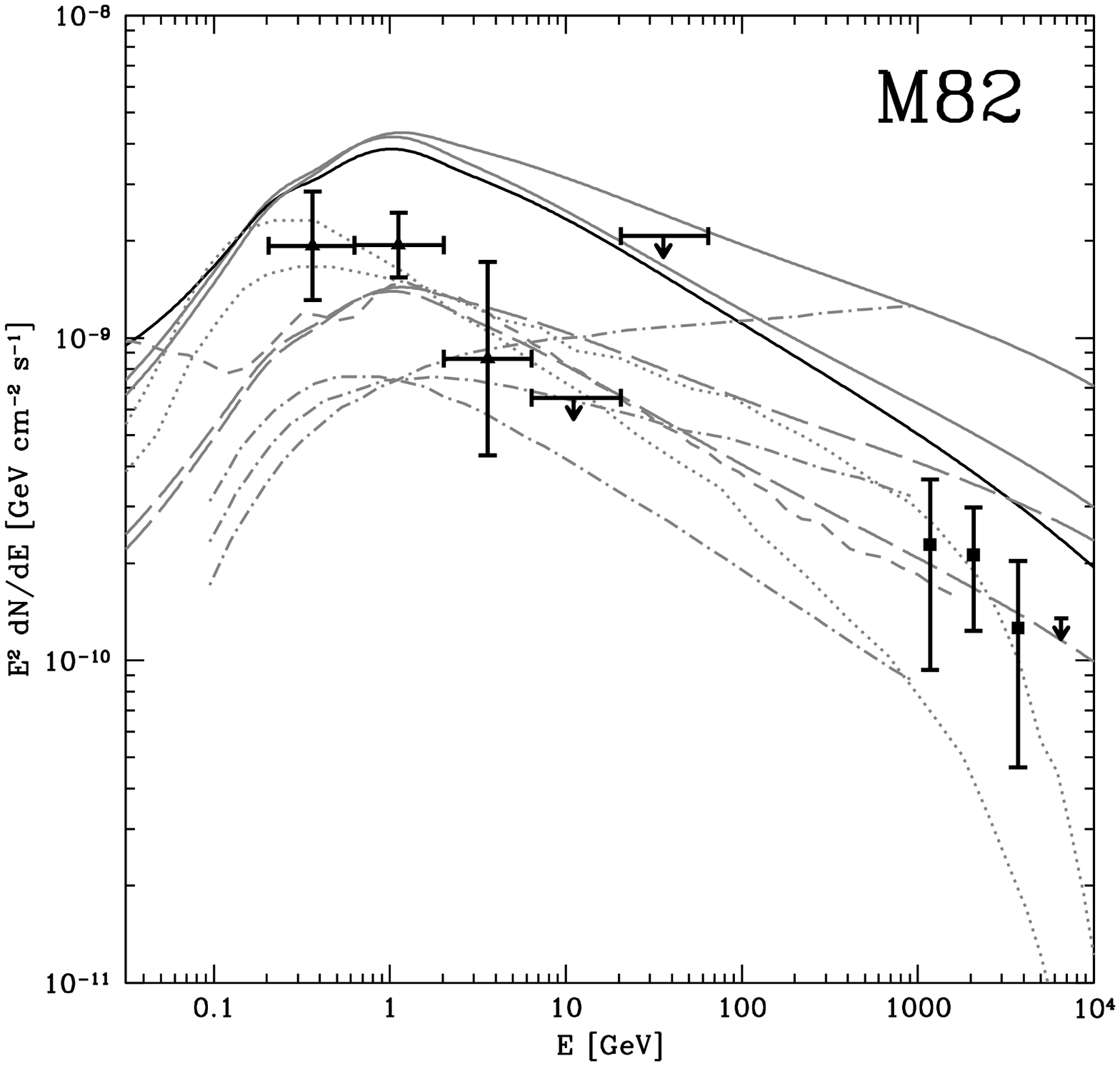}\includegraphics[width=9cm]{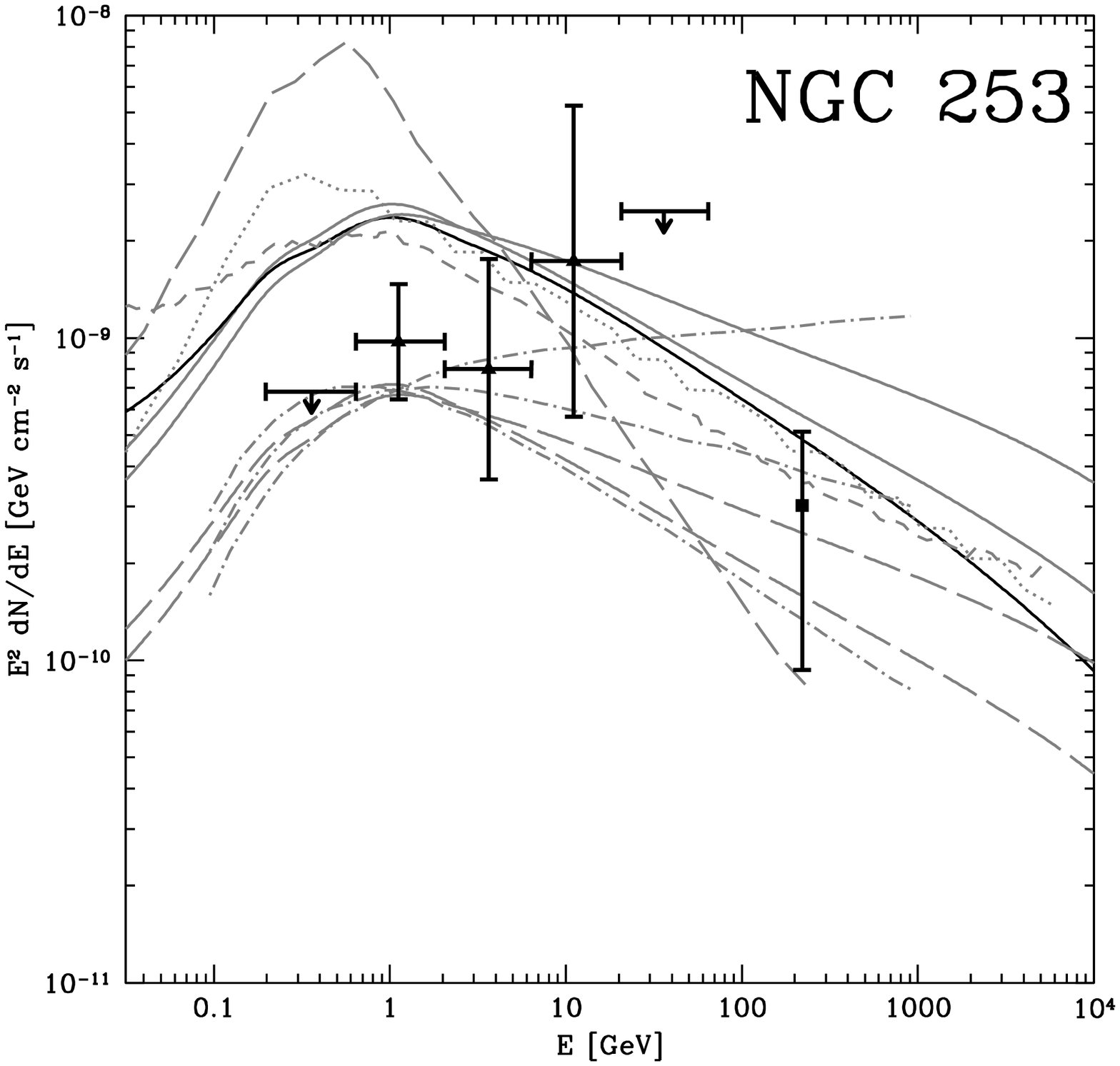}}
\figcaption[simple]{The $\gamma$-ray spectra of M82 and NGC 253 from \emph{Fermi} (\emph{solid triangles}: \citet{Abdo10a}; \emph{open triangles}: analysis in Appendix~\ref{sec:FermiAnalysis}) and VERITAS and HESS (\emph{filled squares}).  We show several models: LTQ (unscaled to SFR and using \citet{Kennicutt98} $\Sigma_g$: long-dashed; scaled to \citet{Sanders03} $L_{\rm TIR}$ and \citet{Kennicutt98} $\Sigma_g$: solid, grey; scaled to \citet{Sanders03} $L_{\rm TIR}$ and $\Sigma_g$ in Table~\ref{table:flux_starburst}: solid, black), \citet{deCeaDelPozo09} and \citet{Domingo05} for M82 and NGC 253 respectively (dotted), \citet{Persic08} for M82 and \citet{Rephaeli09} for NGC 253 (dashed), \citet{Paglione96} for NGC 253 (long-dashed), and TQW ($p = 2.0, 2.2, 2.4$, dash-dotted).  We plot $E^2$ times the differential flux at each energy.  At VHE energies, the main source of variation in the models is the CR injection slope $p$.  Note that the models of LTQ are proton calorimetric, and TQW explicitly assumes proton calorimetry (see Fig.~\ref{fig:GeVtoTIR}).\label{fig:GammaSpectra}}
\end{figure*}

\citet{Abdo10a} reported the detections of M82 (6.8$\sigma$) and NGC 253 (4.8$\sigma$) with the \emph{Fermi} LAT instrument.  At energies above a few hundred MeV, the $\gamma$-ray spectrum of starburst galaxies is expected to be described by a power law spectrum with differential photon fluxes $N(E) = N_0 (E / E_0)^{-\Gamma}$, where $\Gamma$ ($\sim 2$) is the photon spectral index \citep[e.g.,][]{Paglione96}.  Using the GeV data point as a normalization, we adopt GeV differential fluxes of
\begin{eqnarray}
N_{\rm M82} & \approx & 1.9^{+0.5}_{-0.4} \times 10^{-9} \left(\frac{E}{\GeV}\right)^{-\Gamma} \DiffphFluxUnits\label{eqn:DiffGeVM82}\\
N_{\rm NGC 253} & \approx & 0.9^{+0.4}_{-0.3} \times 10^{-9} \left(\frac{E}{\GeV}\right)^{-\Gamma} \DiffphFluxUnits.
\label{eqn:DiffGeVN253}
\end{eqnarray}
The \citet{Abdo10a} maximum likelihood analysis of the sources find $\Gamma$ of $2.2 \pm 0.2 \pm 0.05$ for M82 and $1.95 \pm 0.4 \pm 0.05$ for NGC 253; however, the \emph{Fermi} photon statistics are insufficient for accurate measurements of $\Gamma$.  Using the best-fit power laws in \citet{Abdo10a} gives GeV normalizations that are $60\% - 70\%$ of those in equations~\ref{eqn:DiffGeVM82}-\ref{eqn:DiffGeVN253}, although these fits are influenced by the nondetections at lower and higher energies.  \emph{Fermi} has not detected either starburst above $\sim 20\ \GeV$, and NGC 253 is undetected below $\sim 500\ \MeV$.  The reported spectra are shown in Figure~\ref{fig:GammaSpectra}.

Assuming a power law spectrum from GeV to VHE energies, we can combine the \emph{Fermi} detections with the HESS and VERITAS measurements to derive the spectral slope over this energy range.  The integrated fluxes from M82 reported by VERITAS (at 1.3 - 3.8 TeV) correspond to GeV-VHE spectral slopes $\Gamma$ of 2.19 - 2.25 \citep{Acciari09}, in excellent agreement with the measured spectral slope from \emph{Fermi}.  The HESS detection of NGC 253 at 220 GeV implies $\Gamma = 2.3$ \citep{Acero09}, steeper than the best-fit photon index from the \emph{Fermi} detections, but within the quoted errors.  We adopt $\Gamma = 2.2$ for M82 and $\Gamma = 2.3$ for NGC 253 throughout the rest of this paper. 

Our values of $\Gamma$ are only appropriate if the spectrum is truly a single power law between GeV and TeV energies.  A spectral bump at GeV energies will cause an underlying pionic power-law spectrum to appear steeper than it really is; conversely, a spectral bump at TeV energies will cause it to appear flatter.  The possibility of a ``TeV excess'' is notable particularly because such an excess is seen in the Milky Way (\citealt{Prodanovic07,Abdo08}; see \S~\ref{sec:TeVExcess}).  

Figure~\ref{fig:GammaSpectra} shows that the $\gamma$-ray spectra of both M82 and NGC 253 are in reasonable agreement with most of the previous detailed model predictions \citep{Domingo05,Persic08,deCeaDelPozo09,Rephaeli09}.  However, these models slightly overpredict both the GeV flux and the $\sim 400\ \MeV$ flux of NGC 253 by a factor of $\sim 2 - 3$.  The \citet{Paglione96} model greatly overestimates the $\le \GeV$ flux by a factor of $\sim 5$ (although it used a higher CR acceleration efficiency), and predicts a much softer spectrum ($\Gamma \approx 2.7$) than observed.  Overall, the general agreement between the theory and observations of $\gamma$-rays from M82 and NGC 253 is encouraging \citep[see also][]{deCeaDelPozo09b}, and suggests that models of other starbursts, particularly Arp 220 \citep[][LTQ]{Torres04}, also predict their $\gamma$-ray fluxes with fidelity (\S\ref{sec:OtherStarbursts}).   

\subsection{M82 and NGC 253 Gamma-Ray Luminosities}

For a $dN/dE = N_0 (E / E_{\rm min})^{-\Gamma}$ $\gamma$-ray spectrum from $E_{\rm min}$ to $E_{\rm max}$, the total luminosity at energies greater than $E_{\rm min}$ is 
\begin{equation}
\label{eqn:LGamma}
L_{\gamma} (\ge E_{\rm min}) = 2.4 \times 10^{39}\ \ergps\ N_{-9} D_{3.5}^2 \beta_{\gamma},
\end{equation}
where $N_{-9}=N_0/(10^{-9}\ \DiffphFluxUnits)$, $D_{3.5}=D/3.5\ \Mpc$, and
\begin{equation} 
\beta_{\gamma} = \left\{ \begin{array}{ll}
	(\Gamma - 2)^{-1} \left[1 - \left(\displaystyle\frac{E_{\rm max}}{E_{\rm min}}\right)^{2 - \Gamma}\right] & (\Gamma \ne 2)\\
\\
	\ln\left(\displaystyle\frac{E_{\rm max}}{E_{\rm min}}\right) & (\Gamma = 2)
	\end{array} \right.\,.
\end{equation}
Typically, $\beta_\gamma \approx 2 - 5$.

The GeV flux reported by \emph{Fermi} for M82 is $N_{-9} = 1.9^{+0.5}_{-0.4}$, and the VERITAS detections imply a $\Gamma \approx 2.2$ spectrum extending to at least $E_{\rm max} = 3.8\ \TeV$.  This corresponds to a $\gtrsim$\,GeV luminosity of 
\begin{equation}
\label{eqn:LM82GeV}
L_{\rm M82} (\ge \GeV) \approx 1.9^{+0.5}_{-0.4} \times 10^{40} D_{3.6}^2 \,\ergps.
\end{equation}
If we interpret the HESS detection of NGC 253 as part of a single $\Gamma \approx 2.3$ power law extending at least to $E_{\rm max} = 220\ \GeV$, and use the $N_{-9} = 0.9^{+0.4}_{-0.3}$ value from \citet{Abdo09}, we find that the GeV luminosity of NGC 253 is
\begin{equation}
\label{eqn:LN253GeV1}
L_{\rm NGC 253} (\ge \GeV) \approx 5.6^{+2.5}_{-1.9} \times 10^{39} D_{3.5}^2 \,\ergps.
\end{equation}
As $E_{\rm max}\rightarrow\infty$, the luminosity increases by only $\sim 25\%$.  In light of the harder \emph{Fermi} best-fit spectrum to the GeV data ($\Gamma = 1.95 \pm 0.4 \pm 0.05$), the VHE emission might be a different spectral component (perhaps pulsar wind nebulae; see \S~\ref{sec:DiscreteSources} and \citealt{Mannheim10}) if the pionic emission falls off between 20 and 200 GeV.  However, a harder spectrum with lower $E_{\rm max}$ still implies a similar $\gamma$-ray luminosity ($\sim 7 - 11 \times 10^{39} D_{3.5}^2\ \ergps$) with those assumptions, and the \emph{Fermi} data are not accurate enough to conclude there is a discrepancy.

\section{Interpretation as Pionic Emission}
\label{sec:Interpretation}

\subsection{Motivation for Proton Calorimetry}
\label{sec:Motivation}
The hypothesis that starbursts are proton calorimeters is motivated by the short pionic energy loss time in their dense interstellar media,
\begin{equation}
\label{eqn:tPion}
t_{\pi} \approx 2 \times 10^{5} \yr \left(\frac{n_{\rm eff}}{250\ \cm^{-3}}\right)^{-1},
\end{equation}
where $n_{\rm eff}$ is the average density of the ISM encountered by CR protons before escape \citep{Mannheim94}, and the mean gas density in the starbursts of M82 and NGC 253 is a few hundred $\cm^{-3}$. 

If $t_{\pi}$ is less than the escape timescale, then the system is a proton calorimeter.  CRs may escape by advection in galactic winds or by diffusion.  The wind advection time is 
\begin{equation}
\label{eqn:tWind}
t_{\rm wind} \approx h/v \approx 2 \times 10^5\ \yr\ h_{100} v_{500}^{-1},
\label{twind}
\end{equation}
for a scale height of $h = 100 h_{100} \pc$ and wind speed of $v = v_{500} \kms$.   We expect the diffusive escape time to go as $t_{\rm diff} (E) = t_0 (E / E_0)^{-1/2}$, where $t_0 = 26\ \Myr$ at $E_0 = 3\ \GeV$ in our Galaxy from radioactive isotopes in cosmic rays \citep[cosmic ray clocks;][]{Connell98,Webber03}.  Little is known about the diffusion escape time in starbursts (for example, \citealt{Domingo05} and \citealt{deCeaDelPozo09} assume $t_{\rm diff}\sim 1 - 10\ \Myr$).  If we assume that CRs stream out of the starbursts at the average Alfven speed $v_A = B / \sqrt{4 \pi \rho}$ \citep{Kulsrud69}, then 
\begin{equation}
\label{eqn:tDiffAlf}
t_{\rm diff} \approx 3.5\ \Myr\ h_{100} B_{200}^{-1} n_{250}^{1/2},
\end{equation}
where $B_{200} = B / (200\ \muGauss)$ and $n_{250} = n / (250\ \cm^{-3})$.

From equations (\ref{eqn:tPion}) -- (\ref{eqn:tDiffAlf}), pionic losses dominate advection losses if $n_{\rm eff} \ga 250\ \cm^{-3} h_{100}^{-1} v_{500}$ and diffusive losses at a few GeV if $n_{\rm eff} \ga 40\ \cm^{-3} B_{200}^{2/3} h_{100}^{-2/3}$, again assuming CRs stream out of the starbursts at the Alfven speed.  We see that the advective and pionic loss times are roughly equal in M82 and NGC 253, implying that inelastic proton-proton collisions are an important loss process for GeV protons, and suggesting proton calorimetry (LW06, TQW).

There are two ways to avoid this conclusion: more efficient escape (shorter $t_{\rm adv}$ or $t_{\rm diff}$), or less efficient pion losses (longer $t_{\pi}$, smaller $n_{\rm eff}$).  Both are possible.  In particular, there may be a fast wind component with $v \approx 1000 - 2000\ \kms$ from M82 \citep{Strickland09}.  In addition, the pionic losses are less efficient if the CRs mainly travel through low density gas.  The ISM in starbursts is clumpy, with most of the volume contained in a low density phases \citep[e.g.,][]{Lord96,Mao00,Rodriguez04,Westmoquette09}.  Indeed, there is $\gamma$-ray and radio evidence that CRs do not penetrate deep into molecular clouds in the Galactic Center, so that $n_{\rm eff} \ll \mean{n}$ and proton calorimetry fails \citep{Crocker10a,Crocker10b}.  

If either advective losses or pionic losses dominate, the CR proton and hadronic $\gamma$-ray spectra should both be relatively hard with $\Gamma \approx 2.0 - 2.4$ for standard injection spectra.  This is because the equilibrium one-zone CR proton spectrum is roughly $N(E) \approx Q(E) \tau(E)$, where $Q(E) \propto E^{-p}$ is the proton injection spectrum with $2.0 \la p \la 2.4$ expected, and $\tau(E)$ is the CR proton lifetime including escape and (catastrophic) pion losses.  Both the advective and pionic lifetimes are roughly energy-independent, so they preserve the hard injection spectrum.  By contrast, in the Milky Way CR proton lifetimes are determined by diffusive escape ($t_{\rm diff} \propto E^{-1/2}$), so the resulting GeV to PeV proton spectra go as $E^{-2.7}$ \citep[e.g.,][]{Ginzburg76}.  The harder $\gamma$-ray spectra of M82 and NGC 253 support either proton calorimetry or strong advective losses.

\subsection{Measuring The Calorimetry Fraction}
\label{sec:CalorimetryFraction}

We define $L_{\pi}$ as the power of the starburst in all pion end-products, including hadronic $\gamma$-rays, neutrinos, and secondary electrons and positrons.  The ratio of $L_\pi$ to the injected CR luminosity ($L_{\rm CR}$) for CR protons with kinetic energy per particle ($K$) above the pion-production threshold ($K_{\rm th}$) is then the ``calorimetry fraction,'' which measures the degree to which starbursts are calorimeters:
\begin{equation}
\label{eqn:timeCalorimetry}
F_{\rm cal} \equiv \frac{L_{\pi}}{L_{\rm CR} (K \ge K_{\rm th})} \approx \frac{t_{\rm life}}{t_{\pi}}.
\label{fcal1}
\end{equation}
The lifetime $t_{\rm life}$ of CR protons with $K > K_{\rm th}$ includes all losses --- pionic, ionization, diffusive, and advective.  If proton calorimetry holds, then $t_{\rm life} \approx t_{\pi}$, $L_{\pi} \approx L_{\rm CR} (K \ge K_{\rm th})$, and $F_{\rm cal}\rightarrow1$.\footnote{Note that equation (\ref{fcal1}) ignores the energy dependence of the losses, aside from the restriction that $K > K_{\rm th}$; ionization losses should be subdominant for $K>K_{\rm th}$ \citep{Torres04}, while both advective and pion losses are roughly independent of energy.  If diffusive losses dominate in starbursts, they may be more effective at higher energy, as in the Galaxy.}
In what follows, we restrict the energy range over which we estimate $F_{\rm cal}$ to be $\ge1$\,GeV.  In
particular, we use observed $\gamma$-rays with energies $\ge1$\,GeV to estimate $L_\pi$ for CRs
with $K\ge1$\,GeV.  While a significant fraction of both the pionic $\gamma$-rays and the CRs have energies below 1\,GeV, leptonic emission is expected to become increasingly important at lower energies, contaminating the estimate of the proton calorimetry fraction.

The total injected CR power $L_{\rm CR}$ likely scales with the star-formation rate \citep[c.f.][]{Abdo10f}, and concomitantly, the supernova rate, $\Gamma_{\rm SN}$.  Assuming that with each supernova, a fraction $\eta^{\prime}$ of its kinetic energy goes to primary CR protons with $K \ge 1\,\GeV$,
\begin{equation}
L_{\rm CR} (\ge \GeV) = 3.2 \times 10^{41} \ergps\ E_{51} \eta^{\prime}_{0.1} \Gamma_{\rm SN,\,0.1}
\end{equation}
where $\eta^{\prime}_{0.1} = \eta^{\prime}/0.1$, $E_{51}$ is the energy of the supernova in $10^{51}\ \erg$, and $\Gamma_{\rm SN,\,0.1}=\Gamma_{\rm SN}/0.1$\,yr$^{-1}$.

Even in the proton calorimetric limit ($F_{\rm cal}\rightarrow1$), the GeV $\gamma$-ray luminosity will be significantly smaller than $L_{\rm CR}$.  First, only $\sim 1/3$ of $L_\pi$ ends up as $\gamma$-rays; $L_{\pi} \approx 3 L_{\gamma}$.  Second, a fraction $\beta_{\pi}$ of the pionic $\gamma$-rays from CR protons with $K\ge$\,GeV will have energies $>$\,GeV.  We calculate $\beta_{\pi}$ using the GALPROP\footnote{GALPROP is available at http://galprop.stanford.edu.} pionic cross sections \citep{Moskalenko98,Strong98,Strong00} based on the work of \citet{Dermer86} \citep[see also][]{Stecker70,Badhwar77,Stephens81} and a $K^{-p}$ spectrum from 1\,GeV to 1\,PeV, and find that it ranges from 0.9 ($p = 2.0$) to 0.5 ($p = 2.5$).  Using these factors,  (see eq.~\ref{eqn:LGamma})
\begin{equation}
F_{\rm cal} \approx 0.023 D_{3.5}^{2} N_{-9} \beta_{\gamma} \beta_{\pi}^{-1} E_{51}^{-1} \eta_{0.1}^{\prime -1} \Gamma_{\rm SN,\,0.1}^{-1}.
\label{Fcal}
\end{equation}

We expect that $\Gamma_{\rm SN}$ is proportional to the luminosity from young massive stars and the star formation rate (SFR).   Because most of the stellar luminosity is converted into infrared light by dust in starbursts, if proton calorimetry holds, then the $\gamma$-ray flux of M82 and NGC 253 should simply be a fraction of the total FIR flux.  Furthermore, if $\eta^{\prime}$ is constant for all starbursts, then this ratio of observed fluxes will be constant in the calorimeter limit, so that starbursts should lie on a \emph{linear} ``FIR-$\gamma$-ray correlation" in analogy with the FIR-radio correlation (TQW).  Observations of normal and starburst galaxies do suggest some kind of correlation between SFR and $L_{\gamma}$, but this correlation increases faster than linearly ($L_{\gamma} (\ge 100\ \MeV) \propto {\rm SFR}^{1.4 \pm 0.3}$; \citealt{Abdo10f}), as expected if escape is more efficient in low luminosity galaxies \citep[c.f.][]{Strong10}.

Following TQW, we assume that SFR is related to the total FIR luminosity ($L_{\rm TIR}[8-1000]$\,$\mu$m) by $L_{\rm TIR}=\epsilon {\rm SFR} c^2$, where $\epsilon$ is an IMF-dependent constant (see, e.g., \citealt{Kennicutt98}).  In the calorimeter limit,  
\begin{equation}
\xi_{\rm GeV-TIR}^{\rm cal} \equiv \frac{1}{\beta_{\pi}}\frac{L_\gamma(\ge {\rm GeV})}{L_{\rm TIR}}
\approx 3.1 \times 10^{-4} E_{51} \eta^{\prime}_{0.1}\Psi_{17}
\label{cal_ratio}
\end{equation}
where $\Psi_{17}=(\Gamma_{\rm SN}/\epsilon)/17$\,M$_\odot^{-1}$ depends very modestly on the star formation history of the galaxy considered; it varies from $\sim15$ to $\sim23$ for continuous star formation over timescales of $3 \times 10^7-10^9$\,yr \citep[TQW]{Leitherer99}.  

We scale the TIR luminosities from \citet{Sanders03} (see Tables \ref{table:flux_normal} \& \ref{table:flux_starburst}) to the same distances as the $\gamma$-ray luminosities in equations (\ref{eqn:LM82GeV}) -- (\ref{eqn:LN253GeV1}).  For M82 ($\beta_{\pi}=0.7$), we find that  
\begin{equation}
F_{\rm cal}^{\rm M82}=\xi_{\rm GeV-TIR}^{\rm M82}/\xi_{\rm GeV-TIR}^{\rm cal}
\approx 0.4 \,(E_{51}\eta^{\prime}_{0.1}\Psi_{17})^{-1}.
\label{M82cal_ratio1}
\end{equation}
For NGC 253, we find that $F_{\rm cal}$ is 
\begin{equation}
F_{\rm cal}^{\rm NGC 253}=\xi_{\rm GeV-TIR}^{\rm NGC 253}/\xi_{\rm GeV-TIR}^{\rm cal}
\approx 0.2 \,(E_{51}\eta^{\prime}_{0.1}\Psi_{17})^{-1}
\label{NGC253cal_ratio2}
\end{equation}
for $\Gamma=2.3$ (to 220 GeV) and $\beta_{\pi} \approx 0.6$.  The flux uncertainties in the \emph{Fermi} and HESS detections of NGC 253 indicate that the uncertainty in $F_{\rm cal}^{\rm NGC 253}$ can be significantly reduced by more {\it Fermi} data.  Note that we have used the entire TIR flux of these galaxies, while the starburst cores (and not the outlying disks) probably dominate the $\gamma$-ray emission \citep[see the treatments of NGC 253 by][]{Domingo05,Rephaeli09}.

Alternatively, one may estimate $F_{\rm cal}$ with a distance-independent supernova rate (e.g., from radio source counts).  In this case, $F_{\rm cal}$ (eq.~\ref{fcal1}) retains its strong distance dependence.  For M82 ($D = 3.6 \,\Mpc$, $N_{-9} = 2$, $\Gamma = p = 2.2$, $\beta_{\rm \pi}=0.7$, $\beta_\gamma=3.96$), 
\begin{equation}
F_{\rm cal}^{\rm M82} \approx 0.3\,(E_{51} \eta^{\prime}_{0.1})^{-1}\, D_{3.6}^2\Gamma_{\rm SN,\,0.1}^{-1},
\label{M82cal1}
\end{equation}
where we have scaled to a value of $\Gamma_{\rm SN}$ typically quoted in the literature.
Similarly, 
\begin{equation}
F_{\rm cal}^{\rm NGC\ 253} \approx 0.1\,\,(E_{51} \eta^{\prime}_{0.1})^{-1}\,D_{3.5}^2 
\Gamma_{\rm SN,\,0.1}^{-1},
\label{NGC253cal2}
\end{equation}
for our adopted NGC 253 spectrum (see eq.~\ref{NGC253cal_ratio2}).

The higher numbers for $F_{\rm cal}$ in equations (\ref{M82cal_ratio1}) \& (\ref{NGC253cal_ratio2}) with respect to equations (\ref{M82cal1}) \& (\ref{NGC253cal2})  are easy to understand.  For $\Psi_{17}=1$, and $L_{\rm TIR}$ gives us $\Gamma_{\rm SN} \approx 0.059\ \yr^{-1}$ and $\Gamma_{\rm SN} \approx 0.049\ \yr^{-1}$ for M82 and NGC 253, respectively.  Thus, the nominal values for $\Gamma_{\rm SN} = 0.1$\,yr$^{-1}$ in equations (\ref{M82cal1}) \& (\ref{NGC253cal2}), while well in the range of supernova rates quoted for both systems (\S\ref{sec:Uncertainties}), are larger than those inferred from the total FIR luminosity by a factor of $\sim 1.7 - 2$.

Note that our estimates of $F_{\rm cal}$ in equations (\ref{NGC253cal_ratio2}) and (\ref{NGC253cal2}) for NGC 253 are still $\sim 2 - 5$ times higher than the HESS estimate of $\sim 0.05$, even though we use similar or higher SN rates.  The main reason for this is that \citet{Acero09} assume a GeV-to-TeV spectral slope of $2.1$, whereas we use a GeV-to-TeV spectral slope of $2.3$, derived from the \emph{Fermi} data.  With this spectral slope \citet{Acero09} effectively underestimate the GeV $\gamma$-ray luminosity of NGC 253 by a factor of $\sim 3$ from $1 - 220$\,GeV.

\subsection{Primary Uncertainties in $F_{\rm cal}$}
\label{sec:Uncertainties}

\noindent \emph{Other $\gamma$-ray sources} -- Any $\gamma$-ray source besides pionic emission from CR protons lowers our estimate for $F_{\rm cal}$.  Although it is in principle possible that other sources dominate, e.g., the TeV emission, it is likely that the GeV emission is in fact pionic.  See \S~\ref{sec:Sources}.\\

\noindent \emph{Other IR sources} -- For our main estimates of $F_{\rm cal}$ in equations~(\ref{M82cal_ratio1}) and (\ref{NGC253cal_ratio2}), we have used the total infrared light of each galaxy to measure the star-formation rates of the $\gamma$-ray emitting starbursts.  However, in NGC 253 only about half of the IR emission comes from the starburst \citep{Melo02}.  It is also possible that cirrus emission from old stars contributes to the observed infrared emission, although we do not expect this to be significant within the starburst itself.  Excluding this additional IR light increases the estimates of $F_{\rm cal}$.\\

\noindent \emph{Acceleration efficiency} -- Higher $\eta^{\prime}$ lowers our estimated $F_{\rm cal}$.  In principle, $\eta^{\prime}$ can be as high as $\sim 1$ \citep{Ellison84,Ellison04}. Efficiencies  $\eta^{\prime} > 1$ are also possible if additional CR power comes from sources other than SNe.  We have scaled the above estimates for $F_{\rm cal}$ using $\eta^{\prime} = 0.1$ based on our work on FIR-radio correlation (LTQ, TQW), which constrains $E_{51} \eta^{\prime}$ to be $\sim 0.1$, depending on the CR proton injection spectrum \citep[see also][]{Torres03,Torres04}.  We emphasize that $\eta^\prime$ is the energy per SN explosion in CR protons with energies $\ge1$\,GeV and does not include low-energy CRs.\\

\noindent \emph{Role of supernovae} -- Although we have assumed in equations~(\ref{M82cal1}) and (\ref{NGC253cal2}) (and implicitly assumed in our definition of $\eta^{\prime}$) that SNe are responsible for all of the CRs, this has not been settled \citep[see][]{Butt09}.  There is evidence now that SN remnants accelerate some CRs \citep{Tavani10}, but other sources may also contribute CRs.  The $\gamma$-ray detections of M82 and NGC 253, combined with the $\gamma$-ray detections of quiescent star-forming galaxies, are evidence that $\gamma$-ray emission scales with star-formation rate \citep{Abdo10f}.  However, other possible sources of CRs include stellar winds \citep{Quataert05}, superbubbles \citep{Higdon05,Seaquist07,Butt08}, pulsars \citep{Arons94,Bednarek97,Bednarek04}, and gamma-ray bursts \citep{Waxman95}, which all presumably scale with star formation rate.  It is also possible that the efficiency of some mechanisms, such as superbubble acceleration, are different in starbursts.\\

\noindent \emph{Supernova rates} -- Even if SNe are responsible for CR acceleration, the SN rates in M82 and NGC 253 are highly uncertain.  Estimates of $\Gamma_{\rm SN}$ come from stellar population fitting \citep{Forster03}, line emission \citep{Bregman00,Alonso03}, FIR emission \citep{Mattila01}, comparison of radio sources with models of SN remnants \citep{vanBuren94}, and direct searches for SNe \citep{Mannucci03}.  Methods based on the bolometric emission are complicated by the star-formation history, potential IMF variations in starbursts, including the high-mass ($\gtrsim8$\,M$_\odot$) slope, the shape of the IMF below $\sim1$\,M$_\odot$, and the transition mass between stars that do and do not produce SNe.  Each of these numbers can affect $\Psi$ in equation (\ref{cal_ratio}), although we do not expect large variations from galaxy to galaxy \citep[TQW;][]{Persic10}.  

Methods that use direct searches for SNe or their remnants are more uncertain, and complicated by biases.  For example, many of the radio sources identified as SN remnants in M82 and NGC 253 may be compact HII regions \citep{Seaquist07}.  Methods based on the expansion speed of SN remnants may be complicated by different physical conditions in starbursts \citep{Chevalier01}.  At present, only two confirmed SNe have been observed in M82  (SN 2004am: \citealt{Singer04,Mattila04}; SN 2008iz: \citealt{Brunthaler09b,Brunthaler10}), along with three radio transients over the past $\sim 30\ \yr$ that may be radio SNe \citep{Kronberg85,Muxlow94,Muxlow10}.  In NGC 253, SN 1940e occured $53\farcs$ ($0.9 D_{3.5}\ \kpc$) from the galaxy's center, outside the starburst itself \citep{Kowal71}.  No radio SN was observed in NGC 253 over 17 years of observations, but the implied 95\% confidence limits on SN rate is weak \citep[$\la 2.4\ \yr^{-1}$;][]{Lenc06}.

Overall, $\Gamma_{\rm SN}$ reported in the literature for M82 and NGC 253 span an order of magnitude, from $0.03\ \yr^{-1}$ to $0.3\ \yr^{-1}$.  Early estimates were very high, with $\Gamma_{\rm SN}\approx0.3\ \yr^{-1}$ \citep{Rieke80}.  More recent estimates have revised $\Gamma_{\rm SN}$ downward to $\sim 0.1\ \yr^{-1}$ (\citealt{Muxlow94,Huang94,vanBuren94,Bregman00,Forster01,Mattila01,Lenc06,Fenech08,Fenech10}).  Nonetheless, $\Gamma_{\rm SN}$ remains uncertain at the factor of $\sim 2 - 3$ level, and SN rates down to $0.02\ \yr^{-1}$ are possible for both systems \citep{Forster01,Colina92,Engelbracht98}.\\

\noindent \emph{Galaxy distances} -- The estimates of $F_{\rm cal}$ in equations~(\ref{M82cal_ratio1}) and (\ref{NGC253cal_ratio2}) do not depend on distance, because they depend only on the ratio of TIR and $\gamma$-ray fluxes.  

Equations~(\ref{M82cal1}) and (\ref{NGC253cal2}) instead assume $\Gamma_{\rm SN}$, so that the distances do matter.  Models of the $\gamma$-ray emission also often assume some distance and supernova rate (e.g., \citealt{Domingo05}, LTQ).  The estimated distances to NGC 253 vary, from less than 2.3 Mpc \citep{Davidge90} to 3.9 Mpc \citep{Karachentsev03}.\footnote{Other estimates include $2.5 - 2.7$\,Mpc from \citet{deVaucouleurs78}, 2.6 Mpc from \citet{Puche88}, $2.9 \pm 0.5$ Mpc from \citet{Blecha86}, 3.3 Mpc from \citet{Mouhcine05}, and $3.5 \pm 0.2$ Mpc from \citet{Rekola05}.  A distance of 2.5 Mpc to NGC 253 is commonly quoted \citep[e.g.,][]{Mauersberger96a}, and is used in the \citet{Domingo05} and \citet{Rephaeli09} models of NGC 253; the HESS analysis similarly used 2.6 Mpc \citep{Acero09}.  TQW and LTQ used 3.5 Mpc, based on the Hubble Law.}  This range amounts to an uncertainty of a factor of $\sim 4$ in the $\gamma$-ray luminosity of NGC 253.   Similarly, distances typically quoted for M82 include $3.3$ Mpc \citep{Freedman88}, $3.6 \pm 0.3$ Mpc from Cepheids in M81 \citep{Freedman94}, and $3.9 \pm 0.6$ Mpc from the red giant branch \citep{Sakai99}, amounting to a $\sim 40 \%$ uncertainty in the luminosity of M82. 

\begin{figure*}
\centerline{\includegraphics[width=14cm]{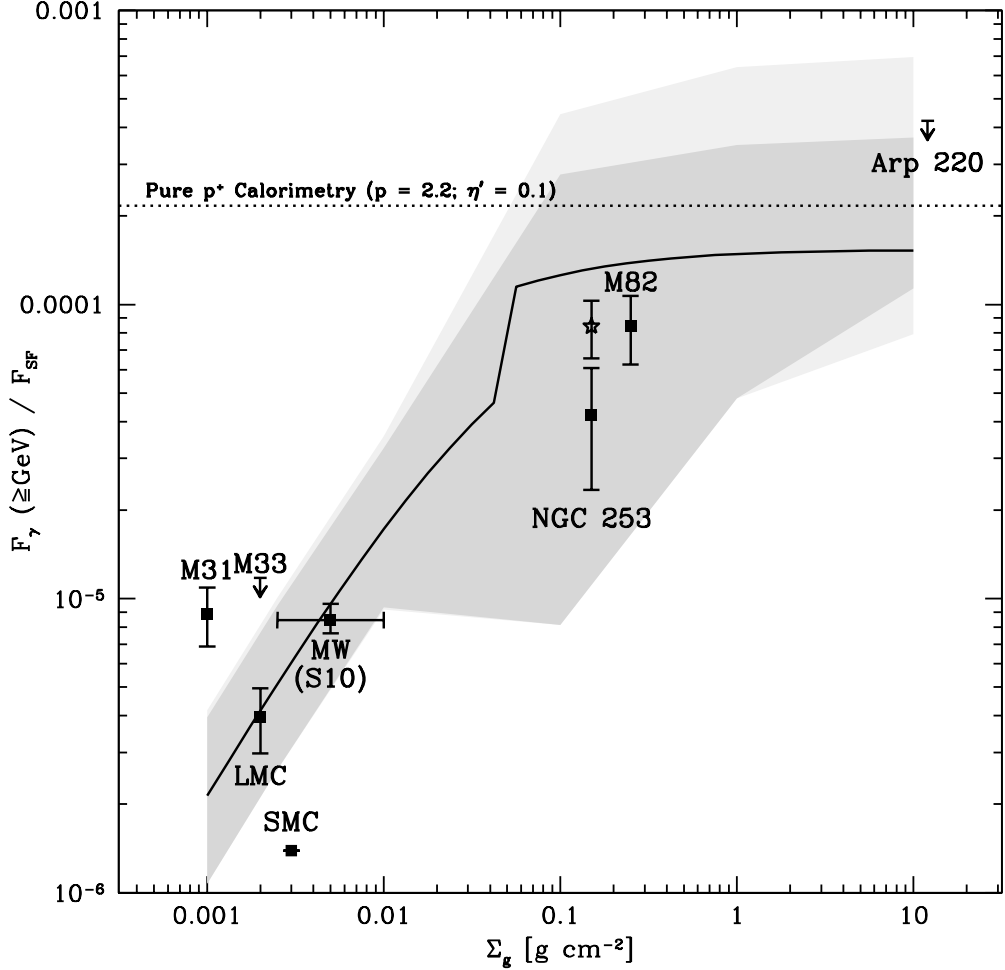}}
\figcaption[figure]{The ratio of the total pionic $\gamma$-ray flux at energies $\ge$\,GeV to the total luminosity from star formation for galaxies with $\gamma$-ray detections or upper limits (for Arp 220, see Appendix \ref{sec:FermiAnalysis}). See Tables \ref{table:flux_normal} \& \ref{table:flux_starburst}. The dashed lines are the calorimetric expectation from equation (\ref{cal_ratio}), scaled to $\eta^{\prime} = 0.05$ and using $\beta_{\pi} = 0.7$ ($p = 2.2$). The solid line is the predicted ratio for pionic $\gamma$-rays in the fiducial model of LTQ based on the Schmidt Law of star formation and the linearity of the FIR-radio correlation.  The model becomes calorimetric at high surface densities where the curve flattens.  The kink at $\Sigma_g = 0.05~\gcm2$ is a result of the scale height changing from $1\ \kpc$ for normal galaxies to $100\ \pc$ for starbursts.  Shading indicates the predicted ratios in all successful models of LTQ (darker: $p = 2.2$ only; lighter: all $p$) for pionic $\gamma$-rays.  The open star represents the ratio for NGC 253's starburst core, if it has one half the TIR luminosity of the entire galaxy \citep{Melo02}.  The $F_{\rm SF}$ of M31, LMC, SMC, and the Milky Way are based on their SFRs \citep{Williams03,Harris04,Harris09,Yin09} and the \citealt{Kennicutt98} conversion factor between $L_{\rm FIR}$ and SFR.  The $F_{\gamma} (\ge \GeV)$ are assumed to be pionic, except for the Milky Way, where the pionic $\gamma$-ray luminosity comes from \citet{Strong10}.  \label{fig:GeVtoTIR}}
\end{figure*}

\subsection{Assessing Proton Calorimetry}
\label{sec:Calorimetry}

The values of $F_{\rm cal}$ derived in Section \ref{sec:CalorimetryFraction} imply efficient proton losses in NGC 253 and M82 compared to the Milky Way.   The $\gamma$-ray data imply both systems have $F_{\rm cal} \approx 0.2 - 0.5$ (compared with $F_{\rm cal} < 0.1$ in the Milky Way based on grammage estimates and modelling; e.g., \citealt{Ginzburg76,GarciaMunoz87,Engelmann90,Jones01,Dogiel02,Strong10}). This suggests that $t_\pi \approx t_{\rm escape}$, even though both galaxies exhibit large-scale galactic winds, and that these systems represent the transition to proton calorimetry.  The case for calorimetry in M82 is stronger than in NGC 253, although $F_{\rm cal}$ is uncertain for NGC 253 by a factor $\sim 2$.  If we only use the TIR flux of the core of NGC 253 (about half the total; \citealt{Melo02}), $F_{\rm cal}$ would be about twice as high, or 0.4 - 0.5 -- the same as M82 (open star in Fig.~\ref{fig:GeVtoTIR}).  Thus, the small $F_{\rm cal}$ for NGC 253 may simply be the result of averaging the $\gamma$-rays from the calorimetric starburst with the non-calorimetric outlying disk.  

Figure \ref{fig:GeVtoTIR} shows the ratio $F_{\gamma} (\ge \GeV) / F_{\rm SF}$ (see Tables \ref{table:flux_normal} \& \ref{table:flux_starburst}), the $\gamma$-ray flux above 1 GeV to the bolometric flux $F_{\rm SF}$ produced by young stars, as a function of gas surface density for NGC 253 and M82, as well as for the LMC \citep{Porter09}, SMC \citep{Abdo10d}, the Galaxy \citep{Strong10}, and M31 \citep{Abdo10f}, together with upper limits on M33 from \citet{Abdo10f} and Arp 220 from our own analysis of the {\it Fermi} data (see Appendix \ref{sec:FermiAnalysis}).  (Note that the plotted ratio $F_{\gamma} (\ge \GeV) / F_{\rm SF}$ does not include $\beta_{\pi}$, since we wish to plot observable quantities.)  The dashed line indicates the calorimetric expectation from equation (\ref{cal_ratio}), scaled to $\eta^{\prime} = 0.1$ and $\beta_{\pi} = 0.7$, for $p = 2.2$.  

The solid line is the prediction of the fiducial model of LTQ, derived by combining constraints from the Schmidt Law of star formation and the observed FIR-radio correlation.  At low gas surface densities, CR protons easily escape, $\gamma$-ray emission is weak and $\xi_{\rm GeV-TIR}^{\rm cal}$ is small.  However, as the gas surface density increases, galaxies become more proton calorimetric and in sufficiently dense starbursts  $\xi_{\rm GeV-TIR}^{\rm cal}$ asymptotes (eq.~\ref{cal_ratio}).  For these galaxies we expect a FIR-$\gamma$-ray correlation (TQW).  The discontinuity at $\Sigma_g = 0.05~\gcm2$ is due to the transition in scale height from $h = 1\ \kpc$ (for normal galaxies) to $h = 100\ \pc$ (starbursts) in the LTQ models; in reality, the transition between normal galaxies and starbursts is smoother.  The $\gamma$-ray luminosity of the LMC is consistent with the predictions of LTQ to within a factor of $\sim 2$.  The standard model of LTQ is tuned to reproduce the Milky Way $\gamma$-ray luminosity given in \cite{Strong00} with $\Sigma_g = 0.01~\gcm2$.  A more recent estimate revises the Milky Way pionic $\gamma$-ray luminosity down by a factor of $\sim 3$ (\citealt{Strong10}: S10); on the other hand, the Milky Way gas surface densities compiled in \citet{Yin09} are also $\sim 3$ times lower (the star-formation rate peaks at $\sim 6\ \kpc$, where $\Sigma_g = 0.003~\gcm2$), so that the Galaxy is still fairly close to the LTQ prediction.  The SMC is $\gamma$-ray dim by a factor of $\sim 4$, suggesting that CRs escape much more easily than expected \citep{Abdo10d}.  Similar behavior is indicated for the CR electrons by the radio emission of irregular galaxies \citep{Murphy08}.  However, M31 is surprisingly $\gamma$-ray bright with respect to the LTQ prediction.  Finally, NGC 253 and M82 appear to be somewhat $\gamma$-ray faint compared to LTQ's fiducial model, consistent with equations (\ref{M82cal_ratio1}) and (\ref{NGC253cal_ratio2}).

A detection of the ULIRG Arp 220 at the level specified in Table \ref{table:flux_starburst} would improve our understanding of how $F_{\rm cal}$ evolves with $\Sigma_g$.  With average densities in its nuclear starbursts exceeding $10^{4}\ \cm^3$ \citep{Downes98}, its pionic loss time is less than $10^4\ \yr$ (eq.~\ref{eqn:tPion}), difficult to reach with winds (eq.~\ref{eqn:tWind}).  According to the models of LTQ, Arp 220 should not have a significantly larger $F_{\gamma} (\ge \GeV) / F_{\rm SF}$ than M82 and NGC 253.  If, instead, Arp 220 is much brighter than M82 and NGC 253, that implies that either $\eta^{\prime}$ is much larger in Arp 220, or escape is more efficient in M82 and NGC 253 than in the fiducial model of LTQ.

\section{Other Sources of $\gamma$-Ray Emission}
\label{sec:Sources}

Our analysis in \S~\ref{sec:Interpretation} assumes that the $\gamma$-ray emission is primarily pionic.  Here, we consider other possibilities that could reduce $F_{\rm cal}$.

\subsection{Diffuse Leptonic Emission} 
\label{sec:Leptonic}

Primary and secondary CR electrons and positrons ($e^{\pm}$) also contribute to the  $\gamma$-ray emissivity of M82 and NGC 253 via bremsstrahlung and Inverse Compton (IC) of predominantly dust-reprocessed starlight.  Detailed models of M82 and NGC 253 indicate that these emission processes are sub-dominant for energies more than $200\ \MeV$ \citep{Paglione96,Domingo05,Persic08,deCeaDelPozo09,Rephaeli09,deCeaDelPozo09b}.  Even without detailed modelling, these processes almost certainly do not dominate $L_\gamma$ in the GeV$-$TeV energy range on energetic grounds.  First, more energy goes to $\pi^0\rightarrow2\gamma$ production than to the secondary $e^{\pm}$.  Second, for a typical ratio of total energy injected in primary electrons to protons of 1/50 \citep[e.g.,][]{Warren05}, and even if all the electron energy goes to producing $\gamma$-rays, over 90\% ($F_{\rm cal} \la 3 / 50$) of the protons would have to escape the starburst for the proton contribution to the $\gamma$-ray luminosity to be sub-dominant.   Assuming that none of the protons interact at all, the overall energy budget of the observed $\gamma$-ray emission implies that the efficiency of primary electron acceleration in M82 and NGC 253 would have to be $\sim 10$ times higher than inferred in the Galaxy.  Finally, the magnetic energy density ($B \approx 200\ \muGauss$ implying $U_B \approx 1000\ \eV\ \cm^{-3}$) is predicted to be at least as strong as radiation energy density ($U_{\rm rad} \approx 200 - 1000\ \eV\ \cm^{-3}$; e.g., \citealt{Paglione96,Persic08}) in many models (e.g., \citealt{Condon91,Domingo05}, TQW, \citealt{Persic08,deCeaDelPozo09,Rephaeli09},LTQ).  Thus, synchrotron losses will increase the energetics requirements even further.  

However, in specific energy ranges leptonic emission can dominate the $\gamma$-ray emissivity.  Bremsstrahlung and IC emission probably make up most of the $\gamma$-ray emission below 100 MeV \citep[e.g.,][LTQ]{Paglione96,Domingo05,Persic08,deCeaDelPozo09,Rephaeli09,deCeaDelPozo09b}.  Bremsstrahlung falls off steeply with energy, and is unimportant above $\sim \GeV$.  The IC spectrum is complicated by the shape of the input photon spectrum.  Given that the photon SED of starbursts is dominated by the FIR, for a CR $e^{\pm}$ injection spectrum $Q(E) \propto E^{-p}$, we expect the IC photon index to be $\Gamma_{\rm IC} \approx p/2 + 1$ at $\sim \TeV$ energies \citep{Rybicki79}.\footnote{Electrons at these high energies, far greater than those observed at GHz frequencies, are cooled almost entirely by IC and synchrotron.}   Whether pionic or advective losses dominate, the pionic emission will have a photon index $\Gamma_{\pi} \approx p$; thus, IC can dominate the VHE $\gamma$-ray emission, but only if $p$ is substantially greater than 2.0 in the calorimetric or advective limit.\footnote{For example, if $p = 2.2$, then $\Gamma_{\pi} = 2.2$ and $\Gamma_{\rm IC} = 2.1$; over 1 dex in $\gamma$-ray energy, this amounts to only a $\sim 30 \%$ increase in the ratio of IC to pionic $\gamma$-rays.}  In practice, Klein-Nishina effects will suppress the IC luminosity beyond a cutoff $E_{\rm KN} \approx  18\ \TeV\ (\lambda/80\ \mu m)$ for target photons of wavelength $\lambda$.  For these reasons, IC is unlikely to contribute significantly to the TeV emission unless $p \gtrsim 2.5$.

At present, the \emph{Fermi} detections presented in \citet{Abdo10a} only use photons with energies greater than 200 MeV.  Improving the limits on 100 MeV photons is essential to determining the leptonic contribution, which should begin to dominate at lower energies.  Direct detection of bremsstrahlung and IC from electrons and positrons would have strong implications for the synchrotron radio emission, and could test the ``high-$\Sigma_g$ conspiracy'' postulated by LTQ to explain the radio emission of starburst galaxies (see \S\ref{sec:FRCImplications}).

\subsection{Discrete $\gamma$-Ray Sources}
\label{sec:DiscreteSources}

Because M82 and NGC 253 are unresolved by \emph{Fermi}, VERITAS, and HESS, the $\gamma$-ray detections include diffuse emission from CRs and emission from discrete sources.\footnote{Emission from AGNs is unlikely to be important, because no variability is observed \citep{Abdo10a} and any AGN luminosity in NGC 253 and M82 is small compared to the star-formation luminosity \citep{Brunthaler09a}.}  The high energy particles in these sources responsible for $\gamma$-rays need not contribute to the general CR population.  Many such sources in the Galaxy are known to be associated with star-formation, including pulsars and a large number of unidentified sources  \citep{Abdo09}, and should be expected in abundance in starbursts.

Because relatively little work has been done on the expected properties of such sources in starbursts, it is unclear if they could dominate the $\gamma$-ray emission from M82 and NGC 253.  A number of star-formation phenomena are known to be TeV sources \citep[as reviewed by, e.g.,][]{Grenier08,Horns08,Hinton09}, including SN remnants, Pulsar Wind Nebulae (PWNe) \citep{Abdo09b}, and possibly star clusters \citep{Aharonian07}.  The observed Galactic TeV sources tend to have hard spectra ($\Gamma \approx 2.2$), similar to the observed GeV-to-TeV spectra of M82 and NGC 253.  

As an example, PWNe have a total energy budget set by the pulsar rotational energy, $E_{\rm rot} \approx 2 \times 10^{50}\ \erg\,P_{0.01}^{-2}$, where $P_{0.01}=P/0.01$\,s is the pulsar spin period at birth, comparable to that injected into CR protons by the SN remnants.  The typical spindown luminosity is $\dot{E}_{\rm rot} \approx 6 \times 10^{39} P_{0.01}^4 B_{12}^2$\,ergs s$^{-1}$, where $B_{12} = B/10^{12}$\,G is the pulsar magnetic field strength, corresponding to a spindown timescale of $\sim 10^3$\,yr.  If all of this energy went into $\gamma$-ray emission in M82 and NGC 253, a pulsar birth rate of $\sim 0.1$\,yr$^{-1}$ could easily power the GeV-TeV emission.  However, Galactic $\gamma$-ray sources like the Crab \citep[e.g.,][]{Albert08}, Geminga \citep{Yuksel09}, and HESS J1825-137 \citep{Aharonian06a} have GeV-TeV luminosities several decades lower than this estimate.  For example, the total $\gamma$-ray luminosity of the Crab ($L_\gamma\approx 10^{35}$\,ergs s$^{-1}$) implies that $\sim 10^4 - 10^5$ such objects would be needed to contribute significantly to the GeV-TeV emission seen from M82 and NGC 253 \citep[see also][]{Mannheim10}.  Given the pulsar birthrate and spindown timescale, this seems unlikely, but we cannot rule out the PWNe in starbursts are much more radiatively efficient in $\gamma$-rays than in the Galaxy, for example, through stronger IC losses.

\subsection{A TeV Excess?}
\label{sec:TeVExcess}
The TeV background of the Milky Way shows a ``TeV excess'' above the expected pionic background \citep{Prodanovic07,Abdo08}.  Whether it is caused by unresolved discrete sources or truly diffuse emission is not known; nor is it known whether it is hadronic or leptonic.   The TeV excess varies with Galactic longitude, being strongest in the Galactic Center and the Cygnus regions \citep{Abdo08}.  The latitude profile of the Galactic TeV $\gamma$-ray emission supports a hadronic explanation for the TeV excess, but leptonic models are not yet excluded \citep{Abdo08}.  

The TeV excess is visible in the Galaxy because the pionic spectrum is steep; effectively, the TeV excess changes $\Gamma$ from 2.7 to 2.6 \citep{Prodanovic07}.  If M82 and NGC 253 have hard pionic $\gamma$-ray spectra at TeV energies, then the $\gamma$-rays from ambient CRs interacting with their ISM will bury any TeV excess.  Furthermore, the simplest explanation for a hadronic TeV excess is that some regions of the Galaxy are denser and more proton calorimetric, and that the TeV excess is simply pionic emission from the normal CR protons.  This effect is observed in molecular clouds located near CR acceleration sites in the Milky Way \citep{Aharonian06b,Albert07,Aharonian08}, but if $F_{\rm cal} \approx 1$, the entire starburst is illuminated this way, and the total pionic $\gamma$-ray luminosity cannot be increased further.  

\section{The GeV-GHz Ratio: A Diagnostic of Electron Cooling and the FIR-Radio Correlation}
\label{sec:FRCImplications}

The observed $\gamma$-rays from M82 and NGC 253 have important implications for the physics of the radio emission of starburst galaxies (see LTQ and references therein).  If pionic, $L_\gamma$ necessarily implies production of secondary $e^{\pm}$s, which produce synchrotron radiation and contribute to the GHz emissivity of the starbursts.  \citet{Rengarajan05}, TQW, and LTQ have all argued that secondary $e^{\pm}$ dominate the synchrotron emission from starbursts.  Detailed models of starburst regions by \citet{Paglione96}, \citet{Torres04}, \citet{Domingo05}, \citet{Persic08}, \citet{deCeaDelPozo09}, and \citet{Rephaeli09} find that secondary $e^{\pm}$ are the majority of $\sim\GeV$ $e^{\pm}$, although the number of primary electrons is still within a factor of a few of the secondaries.

The $\gamma$-ray to radio ratio provides an important constraint on the cooling mechanism of GHz-emitting electrons, if the $\gamma$-rays are pionic.  This ratio can be understood through a simple argument as follows.  Suppose the protons (and secondary $e^{\pm}$) have an $E^{-2}$ spectrum, with equal energy in each log $E$ bin.  The protons lose energy to pions; roughly 2 times as much energy goes into $\gamma$-rays as electrons and positrons.  Furthermore, since $\nu_C \propto E^{1/2}$, the synchrotron emission from each log bin in $e^{\pm}$ energy is spread over 2 log bins in synchrotron frequency.  Therefore, if the $\gamma$-ray emission is dominated by diffuse pionic emission, and if the radio emission is dominated by secondary $e^{\pm}$,
\begin{equation}
\label{eqn:SimpleGeVGHz}
\nu F_{\nu} (E_{\gamma}) = 4 \nu F_{\nu} (E_e) f_{\rm syn},
\end{equation}
where $E_\gamma$ and $E_e$ are the energies of $\gamma$-rays and $e^{\pm}$ respectively from CR protons of the same energy, and $f_{\rm syn} = (t_{\rm syn} / t_{\rm life})^{-1}$ is the fraction of CR $e^{\pm}$ power going into synchrotron.  CR protons produce pionic $\gamma$-rays with $E_{\gamma} \approx 0.1 K_p$ and secondary $e^{\pm}$ with $E_e \approx 0.05 K_p \approx E_{\gamma} / 2$.  Since electrons that emit GHz synchrotron radiation have an energy of $E_{\rm GHz} \approx 560\ \MeV B_{200}^{-1/2} (\nu / \GHz)^{1/2}$, we compare GeV $\gamma$-rays and GHz radio emission. The GHz flux of M82 and the starburst core of NGC 253 are 9 and 3 Jy respectively \citep{Williams09}, while the $\nu F_{\nu} (\GeV)$ fluxes are $3 \times 10^{-12}\,\FUnits$ and $2 \times 10^{-12}\,\FUnits$ respectively.  This implies that $f_{\rm syn}^{-1}$ is $\sim 8$ for M82 and $\sim 17$ for NGC 253, if all of the $\gamma$-ray flux is from their starburst cores.  If we instead consider the total GHz radio emission of NGC 253, 6 Jy, then $f_{\rm syn}^{-1} \approx 8$.  This implies strong non-synchrotron losses, consistent with bremsstrahlung and ionization cooling, and may be a hint that the ``high-$\Sigma_g$ conspiracy'' advocated by LTQ as an explanation for the linear FRC in dense starbursts is operating in M82 and NGC 253.  Note that if some radio emission were from primaries, this would require even greater non-synchrotron losses.  If $B$ is much higher than we suppose, the $e^{\pm}$ accompanying GeV $\gamma$-rays emit at higher frequencies, but since the radio spectrum only goes as $\nu F_{\nu} \propto \nu^{0.3}$ \citep{Klein88,Williams09}, even an order of magnitude increase in $B$ only changes our conclusions by a factor $\sim 2$.  

M82 and NGC 253 appear to have steeper GeV-to-TeV spectral slopes than $p = 2.0$, but the basic conclusions are unchanged with a more careful analysis.    Most of the $e^{\pm}$ emitting at GHz are expected to be secondaries from CR protons, injected with a $E^{-p}$ spectrum.  The power going into $e^{\pm}$ with energy greater than $K_{\rm GHz}$ is $L_{e} (K_e \ge K_{\rm GHz}) = F_{\rm cal} L_{\rm CR,p} (K \ge K_{\rm GHz}) \beta_{\pi,e} (K_{\rm GHz}) f_{e}$, where $f_e$ is the fraction of pionic luminosity going to secondary $e^{\pm}$, and $\beta_{\pi,e}$ is the fraction of secondary $e^{\pm}$ power from CR protons with $K_p \ge K_{\rm GHz}$ that is in $e^{\pm}$ with $K_e \ge K_{\rm GHz}$.  The energy going into $\gamma$-rays with energy greater than a GeV is $L_{\gamma} (\ge \GeV) = F_{\rm cal} L_{\rm CR,p} (K \ge \GeV) \beta_{\pi} (\GeV) f_{\gamma}$.  Finally the radio luminosity is $\nu L_{\nu} = L_{e} (\ge K_{\rm GHz}) (\nu / \GHz)^{1 - p/2} \beta_{\rm syn} f_{\rm sec} f_{\rm syn}$, where $\beta_{\rm syn}$ is a bolometric correction factor and $f_{\rm sec}$ is the fraction of $e^{\pm}$ that are pionic secondaries.  For $p = 2.2$, we have from the GALPROP cross-sections $\beta_{\pi} (\GeV) = 0.7$, $\beta_{\pi,e} (K_{\rm GHz}) = 0.5$, and $\beta_{\rm syn} \approx 0.1$.  Since $f_\gamma \approx 2 f_e$,
\begin{equation}
L_\gamma (\ga \GeV) \approx 30 \nu L_{\nu} f_{\rm sec} f_{\rm syn}^{-1} \left(\frac{\nu}{\GHz}\right)^{p/2-1}.
\end{equation}
This implies that $f_{\rm synch}^{-1}$ is $\sim 5$ for M82 and $\sim 7$ for NGC 253, if all of the $\gamma$-ray flux is from their starburst cores.  It also shows that the simpler estimate in equation~(\ref{eqn:SimpleGeVGHz}) is a useful approximation even when $p \ne 2$.

These estimates are consistent with the idea that most of the radio emission in M82 and NGC 253's starburst are from secondaries undergoing strong non-synchrotron losses.  However the exact values of these ratios are still fairly uncertain.  The main uncertainties are the fraction of $\gamma$-rays from diffuse pionic emission, the fraction of radio emission from primary CR $e^{\pm}$, the fraction of $\gamma$-rays and radio from the starburst cores as opposed to the outlying disk galaxies, and the uncertainties in the $\gamma$-ray fluxes. 

Bremmstrahlung and ionization can easily provide these extra losses.  For $e^{\pm}$ radiating synchrotron at $\nu = \nu_{\rm GHz} \GHz$, the densities when the bremsstrahlung and ionization cooling timescales are comparable to the synchrotron cooling timescale ($t_{\rm brems}/t_{\rm synch}$ and $t_{\rm ion}/t_{\rm synch} \le 1$) are $n_{\rm eff} \ga 67\ \cm^{-3}\ \nu_{\rm GHz}^{1/2} B_{200}^{3/2}$ and $n_{\rm eff} \ga 54\ \cm^{-3}\ \nu_{\rm GHz} B_{200}$, respectively \citep[LTQ;][]{Murphy09}.  These losses are also suggested by the somewhat flattened GHz synchrotron radio spectra observed in starbursts ($\alpha \approx 0.7$), whereas pure synchrotron and IC cooling would lead to steep spectra with $\alpha \approx 1$ \citep{Thompson06}.  Detailed models which do include all of these losses regularly do fit the radio spectra of M82 and NGC 253 \citep{Paglione96,Domingo05,Persic08,deCeaDelPozo09,Rephaeli09} as well as the $\gamma$-ray spectra \citep{deCeaDelPozo09b}.   These models can improve our interpretation of the relevant loss mechanisms.  Finally, with more $\gamma$-ray data from \emph{Fermi}, the lower energy leptonic $\gamma$-ray emission may be detected, which would directly constrain the importance of loss processes such as bremsstrahlung.

\section{Implications}
\label{sec:Implications}
A number of previous studies have found implications for $\gamma$-ray bright starbursts.  We now discuss these implications in light of the current detections of M82 and NGC 253.

\subsection{The Detectability of Other Star-Forming Galaxies}
\label{sec:OtherStarbursts}

Our inference of $F_{\rm cal} \approx$ 0.4 and 0.2 in M82 and NGC 253 implies that a number of local star-forming and starburst galaxies should be visible with next-generation TeV $\gamma$-ray telescopes like CTA \citep[e.g.,][]{Knoedlseder10} and with additional \emph{Fermi} data \citep[see also][TQW]{Pavlidou01,Cillis05}.  For a number of galaxies chosen from the IRAS Bright Galaxy Survey \citep{Sanders03}, Tables \ref{table:flux_normal} and \ref{table:flux_starburst} list the distance, total FIR luminosity and flux, estimates of the gas surface density $\Sigma_g$, and three determinations of the $\gamma$-ray flux:
(1)  $F^{\rm cal}_\gamma(\ge {\rm GeV})$, the purely calorimetric prediction from 
equation (\ref{cal_ratio}), assuming $\beta_\pi=0.7$ and $\Psi_{17}=1$, 
(2)  $F^{\rm LTQ}_\gamma(\ge {\rm GeV})$, the prediction from LTQ  given $\Sigma_g$ and scaled to $F_{\rm TIR}$ (solid line, Fig.~\ref{fig:GeVtoTIR}),
and (3) the observed flux $F^{\rm obs}_\gamma(\ge {\rm GeV})$ in cases where
there exists either a detection or an upper limit.
Our predictions for $F^{\rm cal}_\gamma(\ge{\rm GeV})$ use equation (\ref{cal_ratio}) with nominal 
values of $\beta_{\pi}=0.7$ ($p = 2.2$) and $\eta^{\prime} = 0.1$.  
Tables \ref{table:flux_normal} \& \ref{table:flux_starburst} provides a useful guide to the detectability of all local galaxies to the extent that their TIR and $\gamma$-ray light is dominated by star formation.

\renewcommand{\thefootnote}{\alph{footnote}}
\begin{deluxetable*}{lcccccccc}
\tablecaption{Non-Calorimetric Galaxies: Predicted \& Observed Gamma-Ray Fluxes\label{table:flux_normal}}
\tablehead{ &  &  &  & & \colhead{Predicted} & \colhead{Predicted} & & \\ &  &  &  & \colhead{Calculated} & \colhead{Calorimetric}  & \colhead{LTQ} & \colhead{Observed} & \\ \colhead{Name} & \colhead{$D$\tablenotemark{a}} & \colhead{$L_{\rm TIR}$\tablenotemark{b}} & \colhead{SFR\tablenotemark{c}} & \colhead{$F_{\rm SF}$\tablenotemark{d}} & \colhead{$F^{\rm cal}_\gamma(\ge{\rm GeV})$\tablenotemark{e}} & \colhead{$F^{\rm LTQ}_\gamma(\ge{\rm GeV})$\tablenotemark{f}} & \colhead{$F_\gamma(\ge{\rm GeV})$\tablenotemark{g}} & \colhead{$\Sigma_g$\tablenotemark{h}} \\ & \colhead{\,(Mpc)\,} & \colhead{\,$\log_{10}[L_\odot]$\,} & \colhead{($\Msun \yr^{-1}$)} & \colhead{\,(ergs cm$^{-2}$ s$^{-1}$)\,} & \colhead{\,(ergs cm$^{-2}$ s$^{-1}$)\,} & \colhead{\,(ergs cm$^{-2}$ s$^{-1}$)} & \colhead{\,(ergs cm$^{-2}$ s$^{-1}$)} & \colhead{\,(g cm$^{-2}$)}}
\startdata
LMC & 0.05  & 8.83  & 0.2\tablenotemark{i} & $1.47 \times 10^{-5}$ & $3.2 \times 10^{-9}$\tablenotemark{i} & $6.2 \times 10^{-11}$\tablenotemark{i} & $5.7 \pm 1.4 \times10^{-11}$\tablenotemark{j} & 0.002\tablenotemark{k} \\
SMC & 0.06  & 7.86  & 0.1\tablenotemark{l} & $5.11 \times 10^{-6}$ & $1.1 \times 10^{-9}$\tablenotemark{l} & $3.1 \times 10^{-11}$\tablenotemark{l} & $1 \times 10^{-11}$\tablenotemark{m} & 0.003\tablenotemark{n} \\ 
M31 (NGC 224) & 0.79  & 9.39  & 1.0\tablenotemark{o} & $2.95 \times 10^{-7}$ & $6.4 \times 10^{-11}$\tablenotemark{o} & $6.3 \times 10^{-13}$\tablenotemark{o} & $2.6 \pm 0.6 \times 10^{-12}$\tablenotemark{p} & 0.001\tablenotemark{q} \\
NGC 598 (M33) & 0.84  & 9.07  & 0.5\tablenotemark{r} & $1.30 \times 10^{-7}$ & $2.8 \times 10^{-11}$ & $5.5 \times 10^{-13}$ & $\la 1.5 \times 10^{-12}$\tablenotemark{s} & 0.002\tablenotemark{t}\\
NGC 6946\tablenotemark{u} & 5.32  & 10.16 & 2.6\tablenotemark{v} & $1.69 \times 10^{-8}$ & $3.7 \times 10^{-12}$ & $1.3 \times 10^{-13}$ & \nodata & 0.004\tablenotemark{t}\\
NGC 5457 (M101) & 6.70  & 10.20 & 1.7\tablenotemark{w} & $6.96 \times 10^{-9}$ & $1.5 \times 10^{-12}$ & $2.9 \times 10^{-14}$ & \nodata & 0.002\tablenotemark{t}\\ 
NGC 5194 (M51)\tablenotemark{x} & 8.63  & 10.42 & 3.6\tablenotemark{u} & $8.89 \times 10^{-9}$ & $1.9 \times 10^{-12}$ & $1.0 \times 10^{-13}$ & $\la 8 \times 10^{-11}$\tablenotemark{y} & 0.006\tablenotemark{t}\\ 
NGC 3031 (M81) & 3.63  &  9.47 & 0.3\tablenotemark{z} & $4.19 \times 10^{-9}$ & $9.1 \times 10^{-13}$ & $1.3 \times 10^{-14}$ & \nodata & 0.0015\tablenotemark{t}\\ 
NGC 3521 & 6.84  &  9.96 & 0.9\tablenotemark{v} & $3.54 \times 10^{-9}$ & $7.7 \times 10^{-13}$ & $2.5 \times 10^{-14}$ & \nodata & 0.0035\tablenotemark{t}\\ 
NGC 5055 & 7.96  & 10.09 & 1.3\tablenotemark{v} & $3.77 \times 10^{-9}$ & $8.2 \times 10^{-13}$ & $2.3 \times 10^{-14}$ & $\la 8 \times 10^{-11}$\tablenotemark{y} & 0.003\tablenotemark{t}\\ 
NGC 7331 & 14.71 & 10.58 & 5\tablenotemark{aa} & $4.25 \times 10^{-9}$ & $9.2 \times 10^{-13}$ & $2.2 \times 10^{-14}$ & $\la 8 \times 10^{-11}$\tablenotemark{y} & 0.0025\tablenotemark{t}
\enddata
\footnotetext[1]{Distances from IRAS BGS unless otherwise noted.}
\footnotetext[2]{TIR luminosities from IRAS BGS unless otherwise noted.}
\footnotetext[3]{The star-formation rate as inferred from the literature.  The TIR flux is likely to be inaccurate as a simple SFR indicator at these luminosities \citep{Bell03}.}
\footnotetext[4]{We calculated the bolometric star-formation flux from the SFR, using $F_{\rm SF} = 3.8 \times 10^{-4} c^2 {\rm SFR} / (4 \pi D^2)$, based on the starburst IR to SFR conversion-factor in \citet{Kennicutt98}.  See cavaets in footnote~c.}
\footnotetext[5]{Pionic gamma-ray flux predicted in the explicitly calorimetric limit: 
$F_{\gamma} (\ge{\rm GeV}) = \beta_{\pi} F_{\rm SF} \times 1.8 \times 10^{-4} 
(E_{51} \eta_{0.05}^\prime \Psi_{17})$, using $\beta_{\pi} = 0.7$ as a fiducial value; see equation \ref{cal_ratio}.}
\footnotetext[6]{Pionic gamma-ray flux predicted by the fiducial model of LTQ, using $F_{\rm SF}$. See solid line in Figure \ref{fig:GeVtoTIR}.  Note that leptonic emission (particularly IC) may dominate at the lowest $\Sigma_g$ and increase the $\gamma$-ray fluxes.}
\footnotetext[7]{Measurement of or upper limit on integrated gamma-ray flux of energies $\ge \GeV$.}
\footnotetext[8]{Gas surface density.  Typical uncertainty in this quantity is $\sim 0.3$ dex.}
\footnotetext[9]{\citet{Harris09} find that a SFR of $0.2\ \Msun\ \yr^{-1}$ in the LMC (as used in Fig.~\ref{fig:GeVtoTIR}.}
\footnotetext[10]{Calculated using \citet{Porter09}, assuming that $\Gamma = 2.7$ above 1 GeV.  Integrating the total emission in Figure 8 of \citet{Abdo10e} gives similar results.}
\footnotetext[11]{Calculated using a total gas mass of $6\times10^8$\,M$_\odot$ \citep{Israel97} and $R_{25}\approx4.9$\,kpc.}
\footnotetext[12]{\citet{Harris04} find an average SFR of $0.1\ \Msun\ \yr^{-1}$ in the SMC (as used in Fig.~\ref{fig:GeVtoTIR})  over the past few Gyr, with occasional bursts of star-formation more recently.}
\footnotetext[13]{Calculated from \citet{Abdo10d}, by integrating the total power above 1 GeV plotted in their Figure 5.}
\footnotetext[14]{Calculated using a total gas mass of $4.5\times10^8$\,M$_\odot$ \citep{Israel97} and $R_{25}\approx3.0$\,kpc.}
\footnotetext[15]{\citet{Williams03} find an average SFR of $1\ \Msun\ \yr^{-1}$ in M31 (as used in Fig.~\ref{fig:GeVtoTIR}).}
\footnotetext[16]{Calculated from the Milky Way-scaled GALPROP model in Fig. 2 of \citet{Abdo10f}.  See also the upper limit in \citet{Blom99}.}
\footnotetext[17]{From \citet{Kennicutt98}.  The peak gas surface density compiled in \citet{Yin09} is comparable.}
\footnotetext[18]{\citet{Gardan07} and references therein find star-formation rates of $0.3 - 0.7\ \Msun\ \yr^{-1}$ in M33.}
\footnotetext[19]{Calculated from \citet{Abdo10f}, scaling from the detection of M31 to the upper limit on M33 for the GALPROP spectral template.}
\footnotetext[20]{From \citet{Kennicutt98}.}
\footnotetext[21]{This system also has a central dense starburst component with $\Sigma_g\approx0.04$\,g cm$^{-2}$ \citep{Kennicutt98}, that may be calorimetric, and amounts to $\sim10$\,\% of the total star formation rate.} 
\footnotetext[22]{Scaled from \citet{Leroy08} to the distance listed here.}
\footnotetext[23]{Calculated from \citet{Kennicutt08} using the \citet{Kennicutt98} H$\alpha$ luminosity to SFR conversion.}
\footnotetext[24]{This system also has a central dense starburst component with $\Sigma_g\approx0.06$\,g cm$^{-2}$ \citep{Kennicutt98}, that may be calorimetric, and amounts to $\sim25$\,\% of the total star formation rate.} 
\footnotetext[25]{EGRET upper limits from \citet{Cillis05}.}
\footnotetext[26]{\citet{Kennicutt08} find an H$\alpha$ luminosity equivalent to $0.5\ \Msun\ \yr^{-1}$, while \citet{Davidge06} find a star-formation rate from 10 to 25 Myr ago (roughly the typical lifetime of GeV CR protons in Milky Way-like galaxies) of $0.1\ \Msun\ \yr^{-1}$.}
\footnotetext[27]{\citet{Thilker07} compare star-formation rates derived through several indicators and find them to be $4.4 - 6.3\ \Msun\ \yr^{-1}$.}
\end{deluxetable*}
\renewcommand{\thefootnote}{\arabic{footnote}}

\renewcommand{\thefootnote}{\alph{footnote}}
\begin{deluxetable*}{lccccccc}
\tablecaption{Possible Calorimetric Galaxies: Predicted, \& Observed Gamma-Ray Fluxes \label{table:flux_starburst}}
\tablehead{ 
&  &  &  & \colhead{Predicted} & \colhead{Predicted} & & \\ 
&  &  &  & \colhead{Calorimetric}  & \colhead{LTQ} & \colhead{Observed} & \\ 
\colhead{Name} & \colhead{$D$\tablenotemark{a}} & \colhead{$L_{\rm TIR}$\tablenotemark{b}} 
& \colhead{$F_{\rm TIR}$\tablenotemark{c}} & \colhead{$F^{\rm cal}_\gamma(\ge{\rm GeV})$\tablenotemark{d}} 
& \colhead{$F^{\rm LTQ}_\gamma(\ge{\rm GeV})$\tablenotemark{e}} & \colhead{$F_\gamma(\ge{\rm GeV})$\tablenotemark{f}} 
& \colhead{$\Sigma_g$\tablenotemark{g}} \\ & \colhead{\,(Mpc)\,} & \colhead{\,$\log_{10}[L_\odot]$\,} 
& \colhead{\,(ergs cm$^{-2}$ s$^{-1}$)\,} & \colhead{\,(ergs cm$^{-2}$ s$^{-1}$)\,} 
& \colhead{\,(ergs cm$^{-2}$ s$^{-1}$)} & \colhead{\,(ergs cm$^{-2}$ s$^{-1}$)} & \colhead{\,(g cm$^{-2}$)}}
\startdata
M82 (NGC 3034) & 3.63  & 10.77 & 
$1.42\times10^{-7}$ & $3.08 \times 10^{-11}$ & $2.0 \times 10^{-11}$ & $1.3\times10^{-11}$ & 0.24\tablenotemark{h}\\
NGC 253        & 3.50\tablenotemark{i} & 10.54\tablenotemark{j} & 
$9.09\times10^{-8}$ & $1.97 \times 10^{-11}$ & $1.2 \times 10^{-11}$ & $6.5 \pm 2.5\times10^{-12}$ & 0.15\tablenotemark{k}\\
NGC 4945       & 3.92  & 10.48 & 
$6.23\times10^{-8}$ & $1.35 \times 10^{-11}$ & $8.5 \times 10^{-12}$ & $9.2\pm3.0\times 10^{-12}$\ \tablenotemark{l} & 0.19\tablenotemark{m}\\
NGC 1068 (M77) & 13.70 & 11.27 & 
$3.15\times10^{-8}$ & $6.84 \times 10^{-12}$ & $1.0 \times 10^{-12}$~\tablenotemark{n} & $3.6 \pm 1.0 \times 10^{-12}$ \tablenotemark{o} & 0.02\tablenotemark{n}\\
NGC 5236 (M83) & 3.60  &  10.10 &   
$3.08\times10^{-8}$ & $6.68 \times 10^{-12}$ & $6.6 \times 10^{-13}$ & $\la 4 \times 10^{-11}$\tablenotemark{p} & 0.01\tablenotemark{q,r}\\
IC 342         & 4.60  &  10.17 &   
$2.22\times10^{-8}$ & $4.82 \times 10^{-12}$ & $7.2 \times 10^{-13}$ & \nodata & 0.02\tablenotemark{q}\\
NGC 2146       & 16.47 &  11.07 &   
$1.37\times10^{-8}$ & $2.97 \times 10^{-12}$ & $1.8 \times 10^{-12}$ & $\la 4 \times 10^{-11}$\tablenotemark{p} & 0.14\tablenotemark{q}\\
NGC 3690/IC 694 &  47.74  &  11.88 &   
$1.06\times10^{-8}$ & $2.30 \times 10^{-12}$ & $1.6 \times 10^{-12}$ & \nodata & 2.6 \tablenotemark{q}\\
NGC 1808 &  12.61  &  10.71 &   
$1.02\times10^{-8}$ & $2.21 \times 10^{-12}$ & $1.3 \times 10^{-12}$ & \nodata & 0.09\tablenotemark{q}\\
NGC 1365 &  17.93  &  11.00 &   
$9.87\times10^{-9}$ & $2.14 \times 10^{-12}$ & $1.2 \times 10^{-12}$ & $\la 8 \times 10^{-11}$\tablenotemark{p} & 0.08\tablenotemark{q}\\
NGC 3256 &  35.35  &  11.56 &   
$9.21\times10^{-9}$ & $2.00 \times 10^{-12}$ & $1.3 \times 10^{-12}$ & \nodata & 0.28\tablenotemark{q}\\
NGC 4631 &   7.73  &  10.22 &   
$8.81\times10^{-9}$ & $1.91 \times 10^{-12}$ & $2.8 \times 10^{-13}$ & $\la 4 \times 10^{-11}$\tablenotemark{p} & 0.02\tablenotemark{q}\\
Arp 220 &  79.90  &  12.21 &   
$8.06\times10^{-9}$ & $1.75 \times 10^{-12}$ & $1.2 \times 10^{-12}$ & $\la 8 \times 10^{-11}$\tablenotemark{p} &10\tablenotemark{s}\\
NGC 891 &   8.57  &  10.27 &   
$8.04\times10^{-9}$ & $1.74 \times 10^{-12}$ & $1.0 \times 10^{-12}$ & \nodata & 0.08\tablenotemark{q}\\
NGC 3627\tablenotemark{t} &  10.04  &  10.38 &   
$7.55\times10^{-9}$ & $1.64 \times 10^{-12}$ & $3.6 \times 10^{-13}$ & $\la 4 \times 10^{-11}$\tablenotemark{p} & 0.04\tablenotemark{q}\\ 
NGC 7552 &  21.44  &  11.03 &   
$7.39\times10^{-9}$ & $1.60 \times 10^{-12}$ & $6.7 \times 10^{-13}$ & \nodata & 0.05\tablenotemark{q}\\
NGC 4736 (M94) &   4.83  &   9.73 &   
$7.30\times10^{-9}$ & $1.58 \times 10^{-12}$ & $3.5 \times 10^{-13}$ & \nodata & 0.04\tablenotemark{q}\\
NGC 2903 &   8.26  &  10.19 &   
$7.20\times10^{-9}$ & $1.56 \times 10^{-12}$ & $9.0 \times 10^{-13}$ & $\la 8 \times 10^{-11}$\tablenotemark{p} & 0.08\tablenotemark{q}\\
ESO 173-G015    & 32.44  & 11.34 & 
$6.59\times10^{-9}$ & $1.43 \times 10^{-12}$ & $6.0 \times 10^{-13}$ & \nodata & 0.05\tablenotemark{u} \\ 
NGC 660 &  12.33  &  10.49 &   
$6.45\times10^{-9}$ & $1.40 \times 10^{-12}$ & $8.1 \times 10^{-13}$ & $\la 8 \times 10^{-11}$\tablenotemark{p} & 0.08\tablenotemark{q}\\
NGC 1097 &  16.80  &  10.71 &   
$5.76\times10^{-9}$ & $1.25 \times 10^{-12}$ & $7.4 \times 10^{-13}$ & \nodata & 0.1\tablenotemark{q}\\
NGC 3628\tablenotemark{v} & 10.04 & 10.25 & 
$5.59\times10^{-9}$ & $1.21 \times 10^{-12}$ & $2.3 \times 10^{-13}$ & $\la 4 \times 10^{-11}$\tablenotemark{p} & 0.03\tablenotemark{w}\\
NGC 3079 &  18.19  &  10.73 &   
$5.15\times10^{-9}$ & $1.12 \times 10^{-12}$ & $7.8 \times 10^{-13}$ & $\la 8 \times 10^{-11}$\tablenotemark{p} & 3.7\tablenotemark{q}
\enddata
\footnotetext[1]{Distances from IRAS BGS unless otherwise noted.}
\footnotetext[2]{TIR luminosities from IRAS BGS unless otherwise noted.}
\footnotetext[3]{TIR flux: $F_{\rm TIR}=L_{\rm TIR}/(4\pi D^2)$.}
\footnotetext[4]{Pionic gamma-ray flux predicted in the explicitly calorimetric limit: $F_{\gamma} (\ge{\rm GeV}) = \beta_{\pi} F_{\rm TIR} \times 1.8 \times 10^{-4} (E_{51} \eta_{0.05}^\prime \Psi_{17})$, using $\beta_{\pi} = 0.7$ as a fiducial value; see equation~\ref{cal_ratio}.}
\footnotetext[5]{Pionic gamma-ray flux predicted by the fiducial model of LTQ.  See solid line Figure \ref{fig:GeVtoTIR}.  In these models, leptonic emission is expected to be relatively minor ($\la 10\%$) when integrated above GeV energies, although it may comprise up to $\sim 25\%$ of the differential emission at 1 GeV.}
\footnotetext[6]{Observed gamma-ray flux for energies $\ge$\,GeV, or upper limit.}
\footnotetext[7]{Gas surface density.  Typical uncertainty in this quantity is $\sim 0.3$ dex.}
\footnotetext[8]{We take $M_g = 2.3 \times 10^8 \Msun$ \citep{Weiss01} and $r = 250$ pc for the $D$ adopted: $\Sigma_g = M_g / \pi r^2$.}
\footnotetext[9]{Adopted distance different than in IRAS BGS (3.1\,Mpc) for consistency with the rest of this paper.}
\footnotetext[10]{TIR luminosity corrected for larger adopted distance.}
\footnotetext[11]{From \citet{Kennicutt98}, but scaled to the CO-H$_2$ conversion factor advocated by \citet{Mauersberger96a}.}
\footnotetext[12]{From the 1FGL source catalog, as announced in \citet{Abdo10b}.  NGC 4945 is a Seyfert galaxy, and the AGN may contribute some $\gamma$-ray flux.}
\footnotetext[13]{Total gas mass within a radius of 12'' ($\sim$227 pc at $D = 3.92\ \Mpc$) is taken as $M_g \approx 1.7 \times 10^8 \Msun$ \citep{Mauersberger96b}.}
\footnotetext[14]{\citet{Schinnerer00} give a gas mass of $M_g \approx 5.7 \times 10^8 \Msun$ within $r\approx1.4$\,kpc, implying $\Sigma_g\approx 0.02\ \gcm2$ for NGC 1068.  However, the gas mass is not uniformly distributed in this region, but is concentrated in spiral arms.  If we instead use $\Sigma_g \approx 0.1\ \gcm2$, we find $F^{\rm LTQ}_{\gamma} (\ge \GeV) = 4.0 \times 10^{-12}$ ergs cm$^{-2}$ s$^{-1}$.}
\footnotetext[15]{Derived from the power law fit to NGC 1068 as found by \citet{Lenain10}.}
\footnotetext[16]{EGRET upper limits from \citet{Cillis05}.}
\footnotetext[17]{From \citet{Kennicutt98}.}
\footnotetext[18]{M83 has a central starburst region with a central surface density of $\Sigma_g \approx 0.07~\gcm2$ and scale radius of $\sim 0.6 \kpc$ \citep{Lundgren04}.}
\footnotetext[19]{From \citet{Downes98}.}
\footnotetext[20]{One member of the Leo Triplet (with NGC 3623 and the starburst NGC 3628).}
\footnotetext[21]{Also known as IRAS 13242-5713. \citet{Negishi01} give diameter of 1.1 armin, corresponding to $\sim 9.45$\,kpc.  Using the \citet{Kennicutt98} relation between FIR luminosity and star formation rate, we derive  $\sim 37.7$\,M$_\odot$ yr$^{-1}$ and a surface density of star formation of $\approx0.54$\,M$_\odot$ yr$^{-1}$ kpc$^{-2}$.  Assuming the  Schmidt Law gives an estimate of the gas surface density of  
$\Sigma_g\approx 0.05$\,g cm$^{-2}$.}
\footnotetext[22]{One member of the Leo Triplet (with NGC 3623 and the starburst NGC 3627).}
\footnotetext[23]{\citet{Israel09} gives $M_g \approx 1.5 \times 10^8$M$_\odot$ in the inner 0.6\,kpc, implying  $\Sigma_g\sim0.03$\,g cm$^{-2}$.  On larger scales,  \citet{Irwin96} derive $M_g \approx 1.7\times10^9$M$_\odot$ in the inner $r\approx1.95$\,kpc, implying  again that $\Sigma_g\sim0.03$\,g cm$^{-2}$.}
\end{deluxetable*}
\renewcommand{\thefootnote}{\arabic{footnote}}

Table \ref{table:flux_normal} provides results for normal star-forming galaxies with low $\Sigma_g$ that are not expected to be calorimetric.  Here, $F^{\rm cal}_\gamma(\ge {\rm GeV})$ provides an upper limit to the pionic $\gamma$-ray flux, and $F^{\rm LTQ}_\gamma(\ge {\rm GeV})$ provides the prediction based on the nominal estimate of $F (\ge \GeV) / F_{\rm SF}$ for $\Sigma_g$ (solid line, Fig.~\ref{fig:GeVtoTIR}) and the star-formation rate listed in the table.  Since emission from old stars and different UV opacities can affect the TIR emission \citep{Bell03}, the $\gamma$-ray predictions should be scaled to known star-formation rates when possible.  Furthermore, leptonic processes may increase the $\gamma$-ray emission significantly, especially in low density galaxies.  Table \ref{table:flux_starburst} gives numbers for the starbursts in the IRAS BGS.  Here, $F^{\rm cal}_\gamma(\ge {\rm GeV})\approx F^{\rm LTQ}_\gamma(\ge {\rm GeV})$.\footnote{Note that NGC 5128 (Cen A) has not been included in either Table.}  Uncertainties in Table~\ref{table:flux_starburst} include those in the gas surface density $\Sigma_g$ of the starbursts and the fraction of TIR light associated with the denser starburst regions as opposed to the surrounding galaxy in the whole.

Note that if they are in fact calorimetric, the starburst/AGN systems NGC 4945 and NGC 1068 in Table \ref{table:flux_starburst} should be the brightest on the sky after M82 and NGC 253 at GeV.  We note that \citet{Abdo10b} very recently announced the $\gamma$-ray detection of NGC 4945 (1FGL J1305.4-4928), with a flux within 50\% of the calorimetric prediction and very near the LTQ prediction; \citet{Lenain10} has also announced the detection of NGC 1068.  However, it is still uncertain how much of the $\gamma$-ray flux from these starbursts is from star-formation, and how much is from the Seyfert nuclei \citep{Lenain10}.

The values in Tables~\ref{table:flux_normal} and \ref{table:flux_starburst} should also give the approximate neutrino luminosities of nearby starbursts, since charged pions decay into neutrinos \citep[e.g.,][]{Stecker79}.  Indeed, models of M82 and NGC 253 predict them to be neutrino sources \citep[e.g.,][]{Persic08,deCeaDelPozo09,Rephaeli09}.  However, NGC 253 and NGC 4945 as Southern objects are not detectable with IceCube (though they may be detectable with KM3NET; \citealt{Katz06}), and M82 is at high declination where IceCube's sensitivity is weakest \citep{Abbasi09}.

\subsection{The Diffuse $\gamma$-ray \& Neutrino Backgrounds \\ from Star Formation}
\label{sec:Backgrounds}

The detections of NGC 253 and M82 at GeV$-$TeV energies together with our determination of $F_{\rm cal}$ in these systems has immediate implications for the diffuse $\gamma$-ray and neutrino backgrounds from star formation, as discussed in LW06 and TQW \citep[see also][]{Pavlidou02,Bhattacharya09}.  Several recent studies present calculations of the star-forming galaxy contribution to the $\gamma$-ray background, using the $\gamma$-ray brightness of nearby normal galaxies and starbursts, finding contributions of order $\sim 10 - 50\%$ \citep[e.g.,][]{Fields10,Makiya11,Stecker10}.  Here we present an updated version of TQW's calculation, based on the IR background and this paper's $F_{\rm cal}$.

The pionic emission from starburst galaxies should be concentrated above 100 MeV, with a power law spectrum above $\sim 1\ \GeV$ to $\PeV$ energies if unabsorbed.  For an acceleration efficiency of $\eta^\prime=0.1$ (see LTQ), for CR protons energies larger $\ge$\,GeV, the total integrated $\gamma$-ray background above 1 GeV is, following equation~(\ref{cal_ratio}) and ignoring absorption and redshift effects,
\begin{equation}
F_\gamma (\ge \GeV) \approx 1.0 \times 10^{-6} \eta^\prime_{0.1} f^{\rm burst}_{0.75} F^{\rm cal}_{0.5} F_{20}^{\rm TIR} \,\,{\rm GeV\,\,s^{-1}\,\,cm^{-2}\,\,sr^{-1}},
\label{total_BG}
\end{equation}
for $\Gamma = 2.2$, where $F_{20}^{\rm TIR}$ is the total diffuse extragalactic TIR background in units of 20 nW m$^{-2}$ sr$^{-1}$, $f^{\rm burst}_{0.75}=f_{\rm burst}/0.75$ is the fraction of the TIR extragalactic background produced by starburst galaxies (see TQW), and $F^{\rm cal}_{0.5} = F_{\rm cal} / 0.5$ is the average calorimetric fraction of these starbursts.  As with individual starbursts, the observable neutrino background must be comparable to the pionic $\gamma$-ray background, and have a similar spectrum (c.f. LW06). 

This estimate for the $\gamma$-ray and neutrino backgrounds from CR protons implies that for $E_{\gamma} \ga 1\ \GeV$ and $\Gamma = 2.2$, $\nu I_\nu({\rm GeV}) \sim 2 \times 10^{-7} (E_{\gamma} / \GeV)^{-0.2} \eta^\prime_{0.1} f^{\rm burst}_{0.75} F^{\rm cal}_{0.5} F_{20}^{\rm TIR}\,\,{\rm GeV\,\,s^{-1}\,\,cm^{-2}\,\,sr^{-1}}$, within a factor of $\sim 2$ of the current observations of the extragalactic $\gamma$-ray background (\citealt{Abdo10c}; see also \citealt{Keshet04}).  This indicates that starburst galaxies can be a major source of the $\gamma$-ray background.  More detailed modelling of redshift evolution are necessary to get the spectral dependence correct; most of the star-formation in the Universe occurs at $z \sim 1$ \citep[e.g.,][]{Hopkins06}, so redshift effects will be significant.  Furthermore, the Universe becomes opaque to $\gamma$-rays with observed energy $\ge 100\ \GeV$ at $z = 1$ ($\ge 50\ \GeV$ at $z = 2$; $\ge 20\ \GeV$ at all reasonable $z$; e.g., \citealt{Gilmore09,Finke10}); the $\gamma$-ray background above this energy will cascade down to lower energies.  The corresponding neutrino background will be affected by redshift but not by opacity.  At energies below 1 GeV, the pionic emission should decline because of the decreasing pion production cross-section, as has been discussed in previous works \citep[e.g.,][]{Prodanovic04,Stecker10}.  

The primary uncertainties are still $f_{\rm burst}$ and $F_{\rm cal}$ for starbursts at high $z$.  Importantly, \citet{Daddi10} recently presented CO luminosities of near-infrared selected BzK galaxies at $z \sim 1.5$.  They derive gas masses for these relatively normal star forming galaxies of $\sim 10^{11}\ \Msun$ and radii of $R \sim 3-6 \ \kpc$.  These numbers imply gas surface densities of $\Sigma_g \approx 0.3 M_{11} R_5^{-2} \gcm2$.  Comparing this with Figure \ref{fig:GeVtoTIR} we expect that BzK galaxies are as calorimetric as M82.   These galaxies are an important contributor to the total star formation budget of the universe in the critical redshift range $z \sim 1-2$, thus strengthening the case for a half-calorimetric background as in equation (\ref{total_BG}).  However, we emphasize that this estimate for the BzK galaxies has significant uncertainties (e.g., the CO-to-H$_2$ conversion factor).

\subsection{The Dynamical Importance of Cosmic Rays in Starbursts}
\label{sec:Feedback}

CRs are dynamically important with respect to gravity in the Galaxy \citep{Boulares90}.  They have recently been claimed to be sub-dominant with respect to gravity in starbursts because of strong pion losses (LTQ).  However, CRs may be important in driving winds in such systems \citep[][]{Socrates08}, as in the Galaxy \citep[][]{Chevalier84,Everett08}. 
The observed $\gamma$-ray emission from M82 and NGC 253 can be converted into a constraint on the product of  $n_{\rm eff}$ (see eq.~\ref{eqn:tPion}) and the energy density of the CRs, $U_{\rm CR}$, and hence inform the question of whether or not CRs are dynamically important in these systems;
\begin{equation}
L_{\rm \pi} (K_{\rm CR} \ge \GeV) \approx U_{\rm CR} V f_{\rm GeV} / t_\pi,
\end{equation}
where $L_{\pi} (K_{\rm CR} \ge \GeV) \approx 3 \beta_{\pi}^{-1} L_{\rm \gamma} (\ge \GeV)$, $V$ is the starburst volume, and $f_{\rm GeV}$ is the fraction of the CR energy density in CRs with energy above 1 GeV.  Taking values for the radius and scale height of $R_{250}=R/250$\,pc and $h_{100}=h/100$\,pc,
\begin{equation}
\label{eqn:UCRneff}
U_{\rm CR} n_{\rm eff} \approx 6100\ \eV\ \cm^{-6} f_{\rm GeV}^{-1} R_{250}^{-2} h_{100}^{-1} D_{3.5}^{2} \beta_{\gamma} \beta_{\pi}^{-1} N_{-9}.
\end{equation}
If we assume $R_{250} = h_{100} = 1$ and that CRs interact with an average density of $\sim 250\ \cm^{-3}$, then $U_{\rm CR}$ is $\sim 300 f_{\rm GeV}^{-1}\ \eV\ \cm^{-3}$ for M82 (comparable with \citet{Acciari09}) and $\sim 100 f_{\rm GeV}^{-1}\ \eV\ \cm^{-3}$ for NGC 253 (extrapolating to 1.3 TeV, $12 \eV\ \cm^{-3}$, about twice that of \citealt{Acero09}; however, they quote $n_{\rm eff} \approx 600\ \cm^{-3}$).  For a $K^{-2.3}$ CR spectrum stretching from 10 MeV to infinity \citep{Torres04}, $f_{\rm GeV} \approx 0.25$, mostly in low ($\ll \GeV$) energy CRs.

There is a degeneracy between $U_{\rm CR}$ and $n_{\rm eff}$: at fixed $L_\gamma$, a small $n_{\rm eff}$ can be accommodated by having a high $U_{\rm CR}$, and vice versa.  The value of $n_{\rm eff}$ is not obvious since the ISM of starbursts is highly turbulent and clumpy, with most of the volume filled with gas that is underdense with respect to the mean density.   The pressure required for hydrostatic balance is $P_{\rm hydro} \approx \pi G\Sigma_g\Sigma_{\rm tot}$, where $\Sigma_{\rm tot}$ is the total surface density and is approximately equal to $\Sigma_g$.\footnote{Although the thermal pressure within M82 is an order of magnitude less than $P_{\rm hydro}$ \citep{Lord96,Smith06}, the turbulent pressure is comparable to $P_{\rm hydro}$ \citep{Smith06}.}  We find that the CR pressure $P_{\rm CR}=U_{\rm CR}/3$ is dynamically unimportant: $P_{\rm CR}/P_{\rm hydro} \approx 0.02 f_{\rm GeV}^{-1}$.  Alternatively, if we assume that $P_{\rm CR}\approx P_{\rm hydro}$, we find that $n_{\rm eff} \approx 0.02 f_{\rm GeV}^{-1} \mean{n} \approx 5 - 20$\,cm$^{-3}$, implying that $t_\pi \approx 5\ \Myr$, approximately 25 times longer than the nominal wind escape timescale (see eq.~\ref{twind}).  This requires far more efficient CR acceleration than $\eta^{\prime} \approx 0.1$ (see \S~\ref{sec:CalorimetryFraction}), which we consider unlikely.  Therefore we conclude that $P_{\rm CR}\ll P_{\rm hydro}$.

\section{Conclusion}
\label{sec:Conclusion}
M82 and NGC 253 have now been detected in GeV and TeV $\gamma$-rays, with fluxes roughly comparable to previous detailed predictions.  We have shown that the observed $\gamma$-ray fluxes imply that a fraction $F_{\rm cal} \approx 0.2 - 0.4$ of the energy injected into high energy CR protons is lost to inelastic collisions (pion production) with protons in the ISM (for $\eta^{\prime} = 0.1$).  However, $F_{\rm cal}$ in the range of 0.1 - 1 can be accommodated with different SNe rates and acceleration efficiencies (see the uncertainties in \S~\ref{sec:Uncertainties} and \S\ref{sec:Sources}).  We find a significantly higher $F_{\rm cal}$ for NGC 253 than \citet{Acero09} because NGC 253 has more GeV emission than they expected.  The uncertainty in $F_{\rm cal}$ will decrease significantly with more observations by {\it Fermi}, HESS, and VERITAS.  

A future test of proton calorimetry in M82 and NGC 253 would be a $\gamma$-ray detection of a ULIRG like Arp 220 \citep[c.f.][]{Torres04}.  Arp 220 is more likely to be a proton calorimeter than M82 and NGC 253, with its extremely high average gas density.  If M82 and NGC 253 are not proton calorimeters but Arp 220 is, the ratio of Arp 220's pionic luminosity to its stellar luminosity will be greater than M82 and NGC 253 -- it will be brighter in $\gamma$-rays than expected (see Figure~\ref{fig:GeVtoTIR}, Tables \ref{table:flux_normal} \& \ref{table:flux_starburst}).  Unfortunately, Arp 220's flux is expected to be challenging to detect with \emph{Fermi}, although upper limits alone may be constraining (as in the Appendix).  Stacking searches of ULIRGs may also prove useful.  

Pionic $\gamma$-ray emission implies secondary $e^{\pm}$ production in these starbursts \citep[c.f.][]{Rengarajan05}; from the GHz to GeV ratio, we found evidence of significant non-synchrotron losses.  This suggests that bremsstrahlung and ionization are important energy loss mechanisms for CR electrons and positrons \citep[c.f.][]{Murphy09}.  This would support the idea presented in \citet{Thompson06} that these losses flatten the GHz radio spectrum of starbursts (\S~\ref{sec:FRCImplications}).  It would also support the ``high-$\Sigma_g$ conspiracy'' suggested by LTQ to explain the linearity of the FIR-radio correlation for starbursts, whereby bremsstrahlung, ionization, and IC losses suppress the synchrotron radio emission of CR electrons in starbursts, but proton calorimetry leads to secondary electrons and positrons that boost the radio emission.

Whatever the underlying physics of $\gamma$-ray production in M82 and NGC 253 is, the high fluxes of these starbursts suggest that other starbursts should also be $\gamma$-ray sources.  We compile our predictions in Tables \ref{table:flux_normal} \& \ref{table:flux_starburst}. Considering that much of the star formation in the universe at high-$z$ is in luminous infrared galaxies \citep[e.g.,][]{Elbaz99,Chary01,PerezGonzalez05,Magnelli09}, starbursts might make up a significant fraction ($\sim 1/2$) of the entire $\gamma$-ray background (e.g., \citealt{Pavlidou02}, TQW, \citealt{Bhattacharya09}; \S\ref{sec:Backgrounds}).  If the hadronic interpretation of the $\gamma$-ray flux holds, the neutrino background should also be large (LW06).

Finally, the conclusion that M82 has $F_{\rm cal} \approx 0.4$ and NGC 253 has $F_{\rm cal} \approx 0.2$ implies that the pion cooling timescale is nearly equal to the wind escape timescale, $\sim 2 \times 10^5$\,yr for these systems.  This, in turn, suggests that the CR protons on average interact with ISM near the mean density.  If this is correct, then the CR pressure is significantly below the pressure needed to support each starburst gravitationally, and CRs are not on average dynamically important deep within the starbursts (\S~\ref{sec:Feedback}). 

\acknowledgements
We would like to thank Matthew Kistler, Boaz Katz, and especially John Beacom for many useful discussions.
B.C.L.~is supported in part by an Elizabeth Clay Howald Presidential Fellowship.  T.A.T.~is supported in part by an Alfred P.~Sloan Fellowship.  This work is funded in part by NASA ATP grant \#NNX10AD01G.  E.Q was supported in part by NASA grant NN06GI68G, the David and Lucile Packard Foundation, and the Miller Institute for Basic Research in Science, University of California Berkeley.  A.L. was supported in part by NSF grant AST-0907890, and by NASA grants NNA09DB30A and NNX08AL43G.  EW's research is partially supported by ISF and AEC grants.  This research has made use of the NASA/IPAC Extragalactic Database (NED) which is operated by the Jet Propulsion Laboratory, California Institute of Technology, under contract with the National Aeronautics and Space Administration.

\appendix
\section{\emph{Fermi} Data Analysis for Arp 220}
\label{sec:FermiAnalysis}

We followed the procedure of \citet{Abdo09}, using the publicly available Fermi data reduction software.  The analysis is reviewed in the available online documentation.\footnote{Located at http://fermi.gsfc.nasa.gov/ssc/data/analysis/documentation/ ; the unbinned likelihood tutorial, which we followed, is specifically at http://fermi.gsfc.nasa.gov/ssc/data/analysis/scitools/likelihood\_tutorial.html.}  We downloaded data from the Fermi LAT data server\footnote{Available at http://fermi.gsfc.nasa.gov/ssc/data/.} for METs of 239557417 to 286813463, a total of 18 months.  Data from within $20^{\circ}$ of each source were downloaded.  We created an exposure cube for the entire sky for this time range.  We first divided the \emph{Fermi} energy range into two broad bands: a low energy bin for $100\ \MeV \le E_{\gamma} \le 1\ \GeV$ and a high energy bin for $1\ \GeV \le E_{\gamma} \le 100\ \GeV$.  The source region had a radius of $10^{\circ}$.  The selection was done with \emph{gtselect}. We then selected high quality events with \emph{gtmktime}.  

Using the exposure cube, we created an exposure map around each source using \emph{gtexpmap}.  Finally, we could perform an unbinned likelihood analysis with \emph{gtlike}.  We modeled all of the sources listed in the 1FGL Fermi source catalog within $15^{\circ}$ of Arp 220, the extragalactic background, diffuse Galactic emission, and Arp 220 itself.  In each energy band, we fit a power law to all of the point sources, including Arp 220; both differential flux and integrated flux were considered for Arp 220.  We used the P6\_V3\_DIFFUSE response function.  The Galactic background was modeled with the gll\_iem\_v02.fit background and the extragalactic background was modeled with isotropic\_iem\_v02.txt, both of which are the default models.  We first found a preliminary fit using DRMNFB, and then used the results of that fit to converge to our final fit with MINUIT.

\begin{deluxetable}{ll}
\tablecaption{\emph{Fermi}-LAT $\gamma$-ray fluxes of Arp 220}
\tablehead{\colhead{Property} & \colhead{Value}}
\startdata
\cutinhead{100 MeV - GeV}
$\Phi_{23}$\tablenotemark{a}         & $(1.4 \pm 0.5) \times 10^{-9}\ \phFluxUnits$\\
$\Gamma_{23}$\tablenotemark{b}    & $1.7 \pm 0.4$\\
$\sqrt{TS_{23}}$\tablenotemark{c} & $0.74$\\
$N (100\ \MeV)$\tablenotemark{d}  & $(16 \pm 8) \times 10^{-9}\ \DiffphFluxUnits$\\
\cutinhead{GeV - 100 GeV}
$\Phi_{35}$\tablenotemark{e}         & $(0.30 \pm 0.05) \times 10^{-9}\ \phFluxUnits$\\
$\Gamma_{35}$\tablenotemark{f}    & \nodata\tablenotemark{g}\\
$\sqrt{TS_{35}}$\tablenotemark{h} & $2.2$\\
$N (1\ \GeV)$\tablenotemark{i}    & $(1.2 \pm 0.7) \times 10^{-9}\ \DiffphFluxUnits$\\
\cutinhead{GeV - 100 GeV; $\Gamma = 2.2$}
$\Phi_{35}$\tablenotemark{e}         & $(0.20 \pm 0.15) \times 10^{-9}\ \phFluxUnits$\\
$\sqrt{TS_{35}}$\tablenotemark{h} & $1.7$\\
$N (1\ \GeV)$\tablenotemark{i}    & $(0.24 \pm 0.17) \times 10^{-9}\ \DiffphFluxUnits$
\enddata
\tablenotetext{a}{Best-fit integrated flux from 100 MeV to 1 GeV.}
\tablenotetext{b}{Best-fit photon index between 100 MeV and 1 GeV.}
\tablenotetext{c}{Test statistic for power law fit between 100 MeV and 1 GeV for Arp 220; the square root is roughly the signficance of detection.}
\tablenotetext{d}{Best-fit differential flux at 100 MeV.}
\tablenotetext{e}{Integrated flux from 1 GeV to 100 GeV.}
\tablenotetext{f}{Best-fit photon index between 1 GeV and 100 GeV.}
\tablenotetext{g}{The best-fit model for Arp 220 has a $\Gamma_{35}$ of 5.0, which was the maximum allowed by our source model file.  This is why the differential flux seems large when the integral flux is small compared to M82 and NGC 253.  This value is almost certainly spurious, considering the predicted faintness of Arp 220. (See Table~\ref{table:flux_starburst}.)}
\tablenotetext{h}{Test statistic for power law fit between 1 GeV and 100 GeV for Arp 220; the square root is roughly the signficance of detection.}
\tablenotetext{i}{Best-fit differential flux at 1 GeV.}
\label{table:Arp220Fermi}
\end{deluxetable}

We did not detect Arp 220, as expected.  Our results are summarized in Table~\ref{table:Arp220Fermi}.  For the high energy band, we considered both models where $\Gamma$ for Arp 220 was allowed to vary, and models where it was forced to 2.2.  In the former, $\Gamma$ always was forced to its maximum value of $5.0$, possibly because of a dearth of high energy photons.  Both variants give similar results for the integrated number of photons above 1 GeV, but in the $\Gamma \to 5$ model, the normalization of the differential flux at 1 GeV is much higher.

\end{document}